\documentclass[10pt,journal,compsoc]{IEEEtran}

\usepackage{amsmath,amsfonts}
\usepackage{algorithmic}
\usepackage{algorithm}
\usepackage{array}

\usepackage{textcomp}
\usepackage{stfloats}
\usepackage{url}
\usepackage{verbatim}
\usepackage{graphicx}

\usepackage{xcolor}
\usepackage{bm}
\usepackage{multirow}
\usepackage{makecell}

\usepackage{caption}
\usepackage{amsfonts,amssymb}
\usepackage{graphicx}
\usepackage{epstopdf}
\usepackage{mathrsfs}
\usepackage{amsmath}
\usepackage{latexsym}
\usepackage{booktabs}
\usepackage{float}
\usepackage{subfigure}
\usepackage{algorithm, algorithmic}
\usepackage{bbm}
\usepackage[amsmath,thmmarks]{ntheorem}
\usepackage{bm}
\usepackage{setspace}

\ifCLASSOPTIONcompsoc
  \usepackage{cite}
\else
  \usepackage{cite}
\fi
\ifCLASSINFOpdf
\else
\fi

\usepackage{xcolor}
\usepackage{xpatch}
 
\makeatletter
\def\changeBibColor#1{%
  \in@{#1}{}
  \ifin@\color{blue}\else\normalcolor\fi
}
 
\xpatchcmd\@bibitem
  {\item}
  {\changeBibColor{#1}\item}
  {}{\fail}
 
\xpatchcmd\@lbibitem
  {\item}
  {\changeBibColor{#2}\item}
  {}{\fail}
\makeatother

\begin{document}
\title{Multi-modal Image and Radio Frequency Fusion for Optimizing Vehicle Positioning  \vspace*{-0em}}

\author{{{Ouwen Huan},  {Tao Luo,} \emph{Senior Member,~IEEE},  {Mingzhe Chen,} \emph{Member,~IEEE}}
\IEEEcompsocitemizethanks{\IEEEcompsocthanksitem O. Huan and T. Luo are with the Beijing Laboratory of Advanced Information Network, Beijing University of Posts and Telecommunications, Beijing, 100876, China. \protect E-mail: \protect\url{ouwenh@bupt.edu.cn}; \protect\url{tluo@bupt.edu.cn}

\IEEEcompsocthanksitem M. Chen is with the Department of Electrical and Computer Engineering and Institute for Data Science and Computing, University of Miami, Coral Gables, FL, 33146, USA. \protect E-mail: \protect\url{mingzhe.chen@miami.edu}}
}

\markboth{}%
{Shell \MakeLowercase{\textit{et al.}}: Bare Advanced Demo of IEEEtran.cls for IEEE Computer Society Journals}


\thispagestyle{empty} 

\IEEEtitleabstractindextext{
\begin{abstract}
In this paper, a multi-modal vehicle positioning framework that jointly localizes vehicles with channel state information (CSI) and images is designed. In particular, we consider an outdoor scenario where each vehicle can communicate with only one BS, and hence, it can upload its estimated CSI to only its associated BS. Each BS is equipped with a set of cameras, such that it can collect a small number of labeled CSI, a large number of unlabeled CSI, and the images taken by cameras. To exploit the unlabeled CSI data and position labels obtained from images, we design an meta-learning based hard expectation-maximization (EM) algorithm. Specifically, since we do not know the corresponding relationship between unlabeled CSI and the multiple vehicle locations in images, we formulate the calculation of the training objective as a minimum matching problem. To reduce the impact of label noises caused by incorrect matching between unlabeled CSI and vehicle locations obtained from images and achieve better convergence, we introduce a weighted loss function on the unlabeled datasets, and study the use of a meta-learning algorithm for computing the weighted loss. Subsequently, the model parameters are updated according to the weighted loss function of unlabeled CSI samples and their matched position labels obtained from images. Simulation results show that the proposed method can reduce the positioning error by up to 61\% compared to a baseline that does not use images and uses only CSI fingerprint for vehicle positioning.
\end{abstract}

}
\maketitle

\section{Introduction}
Vehicle positioning technologies have received significant attentions in both academia and industry due to their important role in autonomous vehicles. 
Current global navigation satellite system (GNSS) based vehicle localization methods suffer from serious performance deterioration in urban environments since satellite signals are severely attenuated or blocked by multi-path propagation \cite{1}. To achieve higher localization accuracy in urban areas, radio frequency (RF) fingerprint based localization is drawing increasing interests \cite{2,45}. Compared to GNSS based positioning methods, RF fingerprint based methods have lower latency and higher localization accuracy in urban areas due to the densely distributed base stations (BSs) or road side units (RSUs). Meanwhile, compared to image-based algorithms \cite{Camera,Camera_x,Camera_y}, the RF based methods can localize users in both line-of-sight (LoS) or non-line-of-sight (NLoS) link dominated scenarios, and are robust to severe lighting and weather conditions. However, using RF fingerprints to localize vehicles still confronts with a number of challenges, such as strong dependence on the amount of labeled training data (i.e. RF fingerprints and the corresponding positions), and low transferability of the model among different scenarios due to the different propagation environments.

Recently, a number of existing works \cite{21,23,24,25,22,2,3,4,27,28} have designed several RF fingerprint based methods for indoor and outdoor positioning. The works in \cite{21,23,24,25,22} formulated positioning as classification problems. In particular, the RF fingerprints in training datasets of these works are collected at some predefined reference locations, and the goal of model training is to correctly classify the location of a user equipment to its nearest reference location. Specifically, the work in \cite{21} presented a deep learning (DL) based positioning system using channel state information (CSI) and designed a greedy learning algorithm to reduce training overhead. The authors in \cite{23} proposed a DL based positioning model that uses both CSI and received signal strenght (RSS) for indoor positioning. In \cite{24}, a DL model based on the ResNet \cite{25} architecture was designed to specially estimate the locations of users using non-line-of-sight (NLoS) transmission links. The works in \cite{22} presented a positioning method that exploits the CSI measurements obtained from unsynchronized access points. The works in \cite{2,3,4,27,28} formulated the positioning problems as coordinate regression problems. In \cite{2}, the authors designed an efficient angle-delay channel amplitude matrix (ADCAM) fingerprint, and then employed a convolutional neural network (CNN) to localize targets. The work in \cite{3} introduced two effective methods to process CSI for positioning. In \cite{4}, an attention-augmented residual CNN with a larger receptive field is presented for indoor localization. The work in \cite{27} investigated the fingerprint based localization aided by reconfigurable intelligent surface (RIS). In this work, the authors proposed a new type of fingerprint named space-time channel response vector (STCRV), and presented a novel residual CNN for target positions estimation. In \cite{28}, the authors employed two fully-connected neural networks (NNs) for localization, in which the first NN was used to roughly classify the target positions, and the second NN is used as the regression module to precisely localize targets. However, all of these works \cite{21,23,24,25,22,2,3,4,27,28} need a large amount of labeled data to train their DL models, such that the high positioning accuracy of the models can be guaranteed. Since the user positions need to be estimated with other methods such as GNSS, a large labeled dataset may not be available especially in some urban areas where GNSS based methods are not effective, which may significantly constrain the localization performances of the DL models.

To address the insufficient labeled data problem in the works \cite{21,23,24,25,22,2,3,4,27,28}, one promising method is to consider the use of multi-modal data such as RF data, light detection and ranging (LiDAR) data, and red green blue (RGB) images that carry rich information about the positions of users. Currently, several existing works \cite{29,30,31,32,6,8,33,34} have studied the use multi-modal data for positioning. 
Specifically, the authors in \cite{29} designed an object detection model for vehicle and blockage localization in images images, and realized blockage prediction using a recurrent neural network (RNN). In \cite{30}, the authors leveraged images to proactively predict dynamic link blockages and hand-off for millimeter Wave (mmWave) systems. In \cite{31} and \cite{32}, the authors employed images to achieve fast and low-overhead mmWave/Terahertz (THz) beam tracking. In \cite{6}, a fusion-based deep learning framework operating on images, LiDAR data, and Global Position System (GPS) data was proposed for optimal beam pair prediction. However, none of these works \cite{29,30,31,32,6} considered to use multi-modal unlabeled data to generate labeled data for training DL models. Therefore, they still need a large-sized labeled dataset for model training. 
In \cite{8}, the authors introduced an image-driven representation method to represent all the received signals using a designed RF image. Then, this RF image was combined with an RGB image captured by the camera for positioning. However, this work did not consider how to find a correct user position that each RF signal corresponds to in the RGB image since each RGB image may capture the locations of multiple users. The work in \cite{33} proposed an unsupervised person re-identification system using both visual data and wireless positioning trajectories under weak scene labeling. In \cite{34}, the authors devised a multi-modal context propagation framework for user localization. The proposed framework contains a recurrent context propagation module that enables position information to be fused between visual data and wireless data, and an unsupervised multi-modal cross-domain matching scheme that utilizes the wireless trajectories to constrain the estimation of pseudo labels of visual data. However, the works in \cite{8,33,34} assumed that the user locations can be directly estimated from RF data, since the focuses of these works are on how to find the correct corresponding relationship between RF fingerprints and the users in images, but not on how to train an RF fingerprint based positioning model with RF and visual data.


The main contribution of this work is a novel multi-modal vehicle positioning framework that enables multiple BSs to jointly utilize images and unlabeled CSI fingerprints which can be collected at a lower cost to train their positioning models, and thus accurately localizing multiple vehicles. The key contributions include:
\begin{itemize}
\item We propose to use the vehicle locations obtained from images as the positioning labels of the unlabeled CSI data. 
Specifically, in the considered vehicle positioning scenario, each vehicle can communicate with only one BS, and hence, it can upload its estimated CSI to only its associated BS. Each BS is equipped with a set of cameras, such that it can collect a small number of labeled CSI, a large number of unlabeled CSI, and the images taken by cameras. However, although the collected images contain location information of multiple vehicles, we do not know the corresponding relationship between the unlabeled CSI data the vehicles captured by images. In consequence, the information of vehicle positions obtained from images cannot be directly used as the labels of unlabeled CSI data to train a positioning model.
\item To solve this problem, we design a hard expectation-maximization (EM) algorithm to train our DL model using unlabeled CSI and images. In particular, since we do not know the corresponding relationship between unlabeled CSI and the multiple vehicles in images, we formulate the calculation of the hard EM surrogate loss as a minimum matching problem \cite{10}. Subsequently, the model parameters are updated according to the minimum matching between unlabeled CSI and position labels obtained from images.  
\item Considering that the incorrect matching between CSI and vehicles in images will result in label noises, we employ a meta-learning based weighting algorithm to improve the robustness of the training process. Specifically, we first modify the training objective of the hard EM algorithm at each iteration by assigning a weight parameter to each pair of unlabeled CSI and vehicle in image. Then, using the meta-learning algorithm, the weight parameter is estimated based on the similarity between the pair of unlabeled CSI and vehicle in image, and the labeled CSI samples in the validation dataset. To further avoid overfitting on the small-sized validation dataset, we design a new rectification method towards the weight parameters solved by meta-learning.
\end{itemize}
Simulation results show that the proposed method can reduce the positioning error by up to 61\% compared to a baseline that does not use images for vehicle positioning.

The rest of this paper is organized as follows. The system model of the considered vehicle positioning system are introduced in Section $\textrm{\uppercase\expandafter{\romannumeral2}}$. Section $\textrm{\uppercase\expandafter{\romannumeral3}}$ introduces the proposed meta-learning based hard EM algorithm which can jointly use images and unlabeled CSI data to train the DL based positioning model. In Section $\textrm{\uppercase\expandafter{\romannumeral4}}$, numerical results are presented and discussed. Finally, conclusions are drawn in Section $\textrm{\uppercase\expandafter{\romannumeral5}}$.

\label{sec:2}

\begin{table}\footnotesize
\newcommand{\tabincell}[2]{\begin{tabular}{@{}#1@{}}#1.3\end{tabular}}
\renewcommand\arraystretch{1.3}
\caption[table]{{List of notations}}
\centering
\begin{tabular}{|c|c|}
\hline 
\!\textbf{Notation}\! \!\!& \textbf{Description}  \\
\hline
$B$ & Number of base stations \\
\hline
$V_b$ & Number of vehicles served by BS $b$ \\
\hline
$\boldsymbol{X}_{b,v}^k$ & \makecell{Pilot symbol sequence transmitted from BS $b$ \\ to vehicle $v$ over subcarrier $k$} \\
\hline
$\boldsymbol{y}_{b,v}^k$ & \makecell{Received symbol sequence of \\ vehicle $v$ over subcarrier $k$} \\
\hline
$\boldsymbol{f}_{b,v}$ & Beamforming vector of transmitted signal \\
\hline
$\boldsymbol{h}_{b,v}^k$ & \makecell{Channel from BS $b$ to vehicle $v$ \\ over subcarrier $k$}\\
\hline
$\boldsymbol{n}_{b,v}^k$ & Additive Gaussian noise\\
\hline
$\boldsymbol{\Sigma}_{b,v}^k$ & Covariance matrix of noise\\
\hline 
$N^{\textrm{P}}$ & Number of symbols in pilot sequence\\
\hline 
$N^{\textrm{B}}$ & Antenna number of ULA\\
\hline 
$N^{\textrm{C}}$ & Number of OFDM sub-carriers\\
\hline 
$\boldsymbol{H}_{b,v}$ & CSI matrix of BS $b$ and vehicle $v$\\
\hline
$\left(o^w,x^w,y^w,z^w\right)$& World coordinate system \\
\hline
$\left(o^i,x^i,y^i\right)$ & Image coordinate system \\
\hline
$\left(o^p,u^p,v^p\right)$ & Pixel coordinate system \\
\hline
$\left[u, v\right]^T$ & Pixel coordinate of a vehicle \\
\hline
$\phi$ & Azimuth angle of vehicle \\
\hline
$\theta$ & Elevation angle of vehicle \\
\hline 
$d^{\textrm{H}}$ & Horizontal distance between vehicle and BS \\
\hline 
$\Delta h$ & height difference between camera and vehicle \\
\hline 
$C$ & Number of cameras equipped at each BS\\
\hline 
$\boldsymbol{I}_{b}^c$ & RGB image captured by camera $c$ at BS $b$\\
\hline 
$\mathcal{I}_b$ & Set of RGB images captured at BS $b$\\
\hline 
$\boldsymbol{p}_{b,i}$ & \makecell{Location coordinate of vehicle $i$ \\ detected by the cameras of BS $b$}\\
\hline 
$\mathcal{P}_b$ & \makecell{Set of vehicle locations obtained \\ from images at BS $b$}\\
\hline 
$\hat{V}_b$ & Number of elements in $\mathcal{P}_b$\\
\hline 
$\boldsymbol{p}_{j}$ & \makecell{Location coordinate of vehicle $j$ \\ from images at all BSs}\\
\hline 
$\mathcal{P}$ & \makecell{Set of vehicle locations obtained \\ from images at all BSs}\\
\hline 
$\hat{V}$ & Number of elements in $\mathcal{P}$\\
\hline 
$\mathcal{D}^{\textrm{M}}$ & Multi-modal dataset of all BSs\\
\hline 
$\mathcal{H}^k$ &  Set $k$ of unlabeled CSI data in $\mathcal{D}^{\textrm{M}}$\\
\hline
$\mathcal{D}_b^{\textrm{L}}$ & Labeled training dataset at BS $b$\\
\hline 
$\mathcal{D}_b^{\textrm{V}}$ & Validation dataset at BS $b$\\
\hline 
$\boldsymbol{\omega}_b$ & \makecell{Parameters of the positioning \\ model at BS $b$}\\
\hline 
$\boldsymbol{\Omega}$ & Matrix of $\boldsymbol{\omega}_b$ of all BSs\\
\hline 
\end{tabular}
\end{table}

\begin{figure}[t]
\centering
\setlength{\belowcaptionskip}{-0.15cm}
\includegraphics[width=8.5cm]{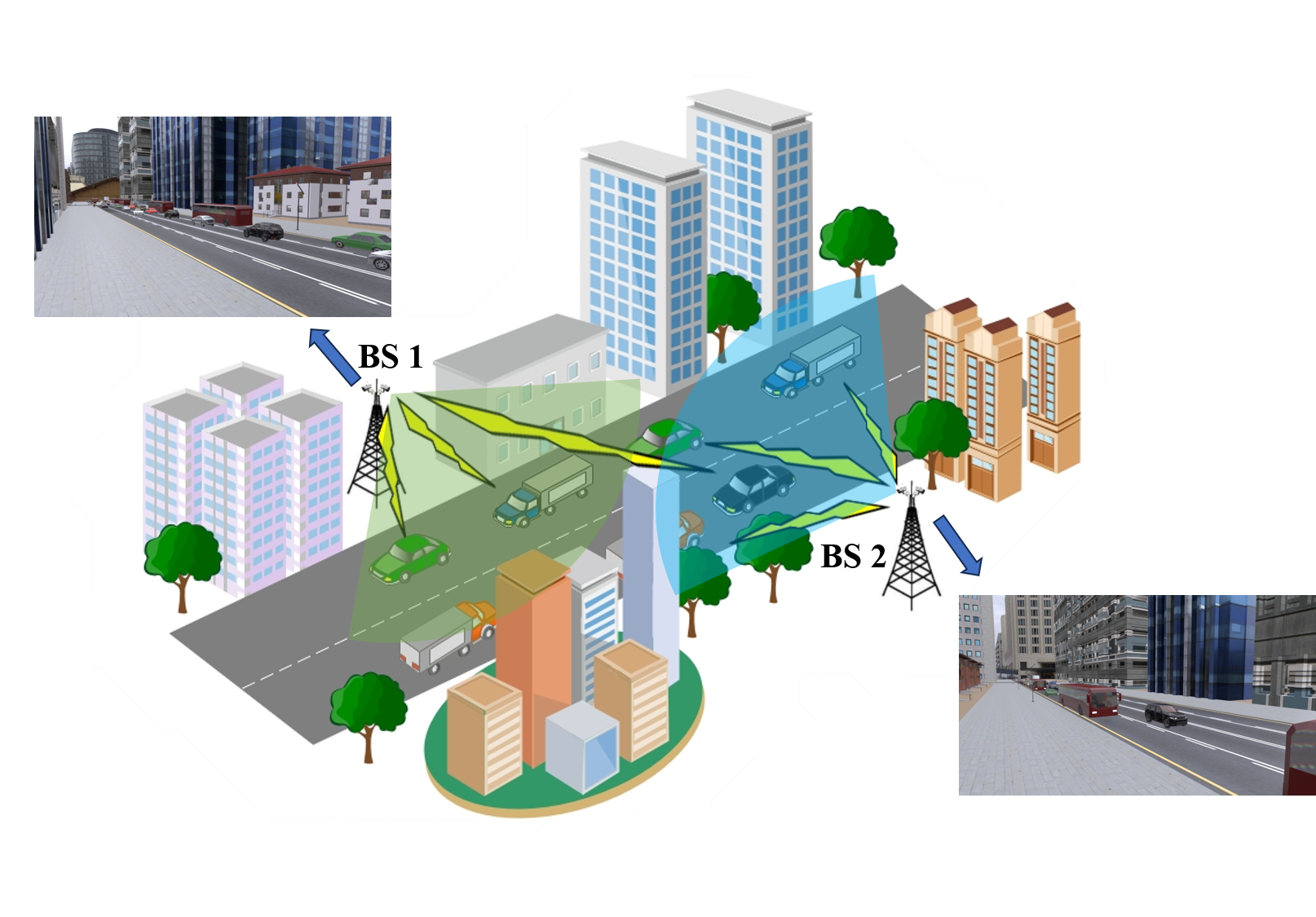}
\centering
\caption{Considered scenario.}
\label{fig1}
\vspace{-0.3cm}
\end{figure}

\section{System Model}
We consider a multi-modal millimeter wave (mmWave) vehicle positioning system where the locations of vehicles are jointly determined by $B$ base stations (BSs) using downlink channel status information (CSI) and red-green-blue (RGB) images as shown in Fig. \ref{fig1}. Specifically, each BS is equipped with a uniform linear array (ULA) with $N^{\textrm{B}}$ antennas, while each vehicle has an omni-directional antenna. To localize a vehicle, the BS will first send a pilot signal to its served vehicles which will send the estimated downlink CSI back. Each BS is equipped with a set of $C$ cameras such that it can use both CSI and images to localize vehicles. We assume that a vehicle can communicate with only one BS at each time slot, and hence, it can upload its estimated CSI to only its associated BS. Next, we first introduce the wireless channel model for explaining the estimation of CSI fingerprints. Then, we introduce the processes of coordinate transformation from pixel coordinates to real-world coordinates, and image processing, in order to clarify how to obtain vehicle location labels from images. Subsequently, we present the datasets and positioning models. Finally, we formulate our multi-modal positioning problem.

\subsection{Wireless Channel Model}
We assume that BS $b$ $\left(b=1,\cdots,B\right)$ is serving $V_{b}$ vehicles in one time slot. The pilot symbol sequence transmitted by BS $b$ to vehicle $v$ $\left(v=1,\cdots,V_{b}\right)$ over subcarrier $k$ is $\boldsymbol{X}_{b,v}^{k}\in\mathbb{C}^{N^{\textrm{P}} \times N^{\textrm{B}}}$, where $N^{\textrm{P}}$ ($N^{\textrm{P}} \textgreater N^{\textrm{B}}$) is the number of symbols in the symbol sequence. Then, the received pilot symbol sequence $\boldsymbol{y}_{b,v}^{k}\in\mathbb{C}^{N^{\textrm{P}} \times1}$ is 
\begin{equation}\label{eq:Communication Model}
\boldsymbol{y}_{b,v}^{k}= \bigg[\boldsymbol{X}_{b,v}^{k} \cdot \big(\boldsymbol{f}_{b,v}\big)^{T}  \bigg] \cdot \boldsymbol{h}_{b,v}^{k} + \boldsymbol{n}_{b,v}^{k},
\end{equation}
where $\boldsymbol{h}_{b,v}^{k}\in\mathbb{C}^{N^{\textrm{B}}\times1}$ is the channel from BS $b$ to the vehicle $v$. $\boldsymbol{f}_{b,v}\in\mathbb{C}^{N^{\textrm{B}} \times 1}$ is the beamforming vector, and $\boldsymbol{n}_{b,v}^{k}$ is the additive noise vector following a zero-mean Gaussian distribution with a covariance matrix $\boldsymbol{\Sigma}_{b,v}^{k}\in \mathbb{R}^{N^{\textrm{P}} \times N^{\textrm{P}}} $. Since both $\boldsymbol{X}_{b,v}^{k}$ and $\boldsymbol{f}_{b,v}$ are known by vehicle $v$, the channel $\boldsymbol{h}_{b,v}^{k}$ can be estimated at vehicle $v$ based on $\boldsymbol{y}_{b,v}^{k}$, and will be transmitted back to BS $b$. The received CSI matrix from vehicle $v$ is
$
\boldsymbol{H}_{b,v}=\left[{\boldsymbol{h}_{b,v}^{1}},{\boldsymbol{h}_{b,v}^{2}},\cdots,{\boldsymbol{h}_{b,v}^{N^{\textrm{C}}}}\right]\in\mathbb{C}^{N^{\textrm{B}}\times {N^{\textrm{C}}}},
$
with $N^{\textrm{C}}$ being the number of valid subcarriers. Then, the set of CSI matrices collected by all BSs at one certain time slot is
$\mathcal{H}=\Big\{\boldsymbol{H}_{b,v}\vert  b \in \left\{1,2,\cdots,B\right\},  v \in \left\{1,2,\cdots,V_{b}\right\}     \Big\}.$


\subsection{Coordinate Models for Image Aided Vehicle Positioning}
\begin{figure}[t]
\centering
\setlength{\belowcaptionskip}{-0.15cm}
\includegraphics[width=8.5cm]{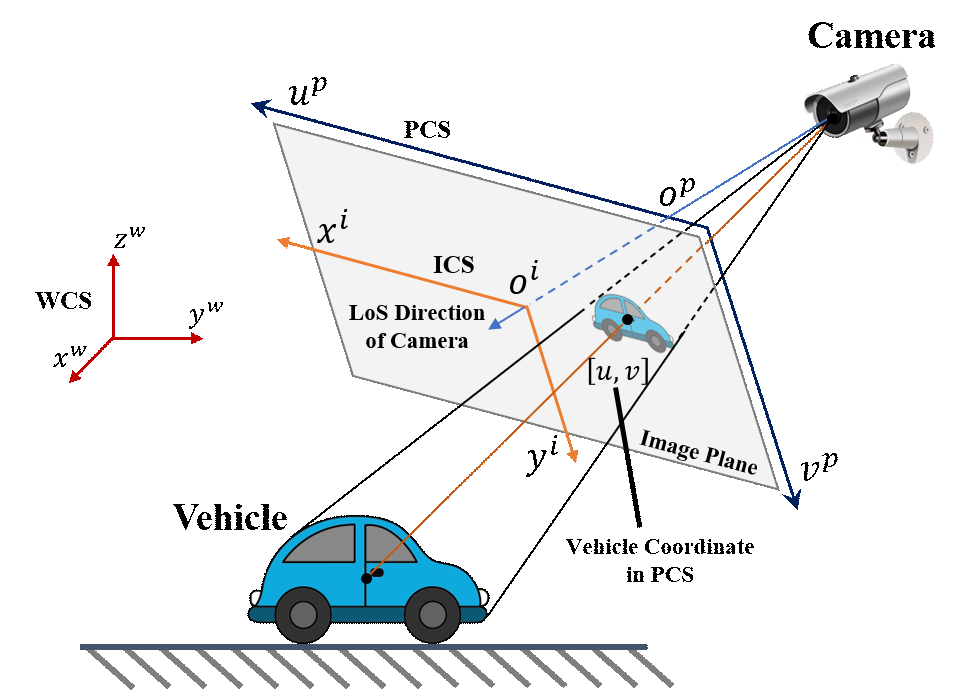}
\centering
\vspace{0.1cm}
\caption{Camera FoV.}
\label{fig2}
\vspace{-0.1cm}
\end{figure}

\begin{figure}[t]
\centering
\setlength{\belowcaptionskip}{-0.15cm}
\includegraphics[width=8.5cm]{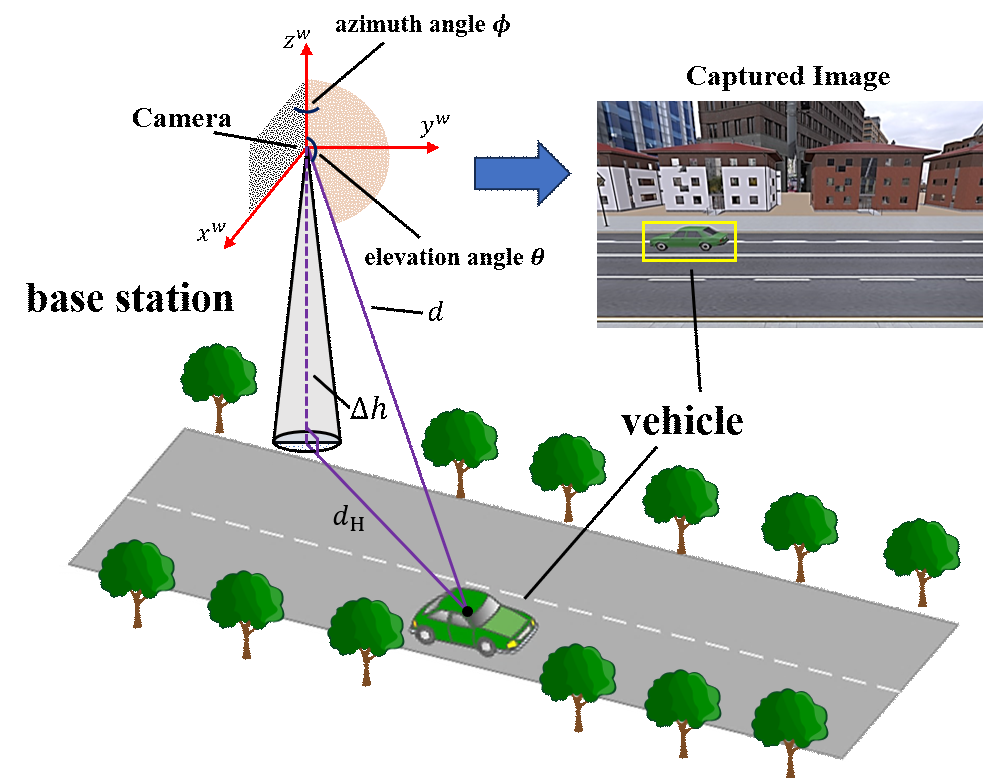}
\centering
\vspace{0.1cm}
\caption{Diagram of vehicle positioning.}
\label{fig3}
\vspace{-0.1cm}
\end{figure}

Since images can only provide pixel coordinates of vehicles, we need to transform these vehicle pixel coordinates to real-world coordinates, such that they can be used for positioning. To explain the coordinate transformation, we consider three coordinate systems as shown in Fig. \ref{fig2}, which are the 3-dimensional (3D) world coordinate system (WCS) $\left(o^w,x^w,y^w,z^w\right)$, the 2-dimensional (2D) image coordinate system (ICS) $\left(o^i,x^i,y^i\right)$ and the 2D pixel coordinate system (PCS) $\left(o^p,u^p,v^p\right)$ \cite{11}. Both the ICS and PCS are on the image plane but have their own coordinate origins (i.e. $o^i \neq o^p$). In particular, we assume that both axis $o^{p} u^{p}$ and axis $o^{i} x^{i}$ are parallel to plane $x^{w} o^{w} y^{w}$. Given the line-of-sight (LoS) direction and viewing angles of a camera, and the width and height of the images, the pixel coordinate $\left[u,v\right]^{\textrm{T}}$ of each vehicle in PCS can be transformed to a polar coordinate $\left[\phi,\theta\right]^{\textrm{T}}$ in the WCS \cite{18}, where $\phi$ and $\theta$ respectively denote the azimuth and elevation angles of the polar coordinate. As shown in Fig \ref{fig3}, the goal of positioning is to estimate the coordinate $\boldsymbol{p}=\left[x,y\right]^{\textrm{T}}=\left[d^{\textrm{H}}\cos{\phi},d^{\textrm{H}}\sin{\phi}\right]^{\textrm{T}}$ of each vehicle in the $x^{w} o^{w} y^{w}$ plane of WCS with $d^{\textrm{H}}$ being the horizontal distance between vehicle and the BS. To estimate $d^{\textrm{H}}$, we first assume that the height difference $\Delta h$ between cameras and each vehicle is roughly the same. Then, given $\Delta h$, the horizontal distance is expressed as $d^{\textrm{H}}={\Delta h}\cdot \tan{\theta}$. Therefore, if we can obtain the pixel coordinate of a vehicle, we can approximately localize this vehicle in 3D world coordinate system through coordinate transformation.

\subsection{Camera Image Processing}
In this section, we introduce how to process the images captured at each BS to obtain vehicle position coordinates from images. Since each BS is equipped with $C$ cameras to capture a set of $C$ images in each time slot, the set of $C$ images captured by BS $b$ at a time slot is
$\mathcal{I}_{b}=\left\{ \boldsymbol{I}_{b}^{c}\vert c=1,2,\cdots,C\right\},$
with $\boldsymbol{I}_{b}^{c}$ being an RGB image captured by camera $c$. All the images have the same dimension of $3\times W\times H$ with $W$ and $H$ being the width and height of the images. Given $\mathcal{I}_{b}$, we first utilize YOLO \cite{12} to detect vehicles from these images in order to obtain their pixel coordinates, such that we can then transform these pixel coordinates of vehicles to their coordinates in WCS. Here, we can also employ other mature object detection models to detect vehicles in images. The set of vehicle position coordinates obtained from all the images in $\mathcal{I}_{b}$ is defined as $\mathcal{P}_{b}=\left\{ \boldsymbol{p}_{b,i}\vert i=1,2,\cdots,\hat{V}_b \right\}$, where $\hat{V}_b$ denotes the number of vehicles detected from images. Note that $\hat{V}_b$ is not certainly equal to $V_b$, since several vehicles served by BS $b$ may not be captured by the cameras due to obstructions or be not in the camera's field of view (FoV). However, the undetected vehicles of BS $b$ may be captured by the cameras of other BSs. To obtain the locations of all vehicles captured by cameras, BS $b$ will share $\mathcal{P}_b$ with other BSs. Then, by integrating $\mathcal{P}_{1},\mathcal{P}_{2},\cdots,\mathcal{P}_{B}$ and removing the duplicate position coordinates, the set of vehicle locations captured by all the images of all BSs at one certain time slot is 
$\mathcal{P}=\left\{\boldsymbol{p}_{j} \vert j=1,2,\cdots,\hat{V}\right\},$
with $\hat{V}$ being the number of vehicles captured by images of all BSs in one time slot.

\subsection{Labeled and Multi-modal Datasets}
Since ground truth vehicle locations are hard to collect due to privacy concerns, we assume that most of the vehicles will upload their CSI to their associated BSs, but only a small number of vehicles will provide both CSI and their locations. As a result, each BS $b$ has a larged-sized unlabeled CSI dataset $\mathcal{D}_b^{\textrm{U}}=\left\{ \boldsymbol{H}_{b,k}\right\}_{k=1}^{N_b^{\textrm{U}}}$ where $\boldsymbol{H}_{b,k}\in\mathbb{C}^{N^{\textrm{B}}\times N^{\textrm{C}}}$ is unlabeled CSI sample $k$, and $N_b^{\textrm{U}}$ denotes the number of unlabeled CSI samples in $\mathcal{D}_b^{\textrm{U}}$, and a small-sized labeled dataset expressed as $\mathcal{D}_b^{\textrm{L}}=\left\{ \boldsymbol{H}_{b,j},\boldsymbol{p}_{b,j}\right\}_{j=1}^{N_b^{\textrm{L}}}$, with $\boldsymbol{H}_{b,j}\in\mathbb{C}^{N^{\textrm{B}}\times N^{\textrm{C}}}$ being CSI sample $j$, $\boldsymbol{p}_{b,j}\in\mathbb{R}^{2 \times 1}$ being the corresponding coordinate of $\boldsymbol{H}_{b,j}$, and $N_b^{\textrm{L}}$ being the number of data samples of labeled dataset $\mathcal{D}_b^{\textrm{L}}$. To create labels for the unlabeled CSI data, one can obtain vehicle locations from RGB images. Therefore, the multi-modal dataset consists of unlabeled CSI and vehicle locations obtained from images can be defined as $\mathcal{D}^{\textrm{M}}=\left\{ \mathcal{H}^{k},\mathcal{P}^{k}\right\}_{k=1}^{{N^{\textrm{M}}}}$, where $N^{\textrm{M}}$ is the number of samples in $\mathcal{D}^{\textrm{M}}$, $\mathcal{H}^{k}$ is the set of CSI collected by all the BSs, and $\mathcal{P}^{k}$ represents the position coordinates obtained from images of all BSs. We assume that the set of vehicle positions obtained from images (i.e. $\mathcal{P}^k$) is a subset of the true labels of the unlabeled CSI samples in $\mathcal{H}^k$. However, we do not know the exact relationship between the unlabeled CSI in $\mathcal{H}^k$ and the location labels in $\mathcal{P}^k$.


\subsection{Positioning Model}
Each BS uses a neural network (NN) with a similar structure to localize its served vehicles. We model the NN of BS $b$ as a mapping from the space of CSI matrices to the space of vehicle locations, which is given by
\begin{equation}\label{eq:Pretraining Dataset}
F_{\boldsymbol{\omega}_b}\left(\cdot\right):\mathbb{C}^{N^{\textrm{B}}\times N^{\textrm{C}}}\rightarrow \mathbb{R}^{2 \times 1},
\end{equation}
where $\boldsymbol{\omega}_b$ denotes the parameter vector of the NN at BS $b$. The parameter matrix of all the NNs is defined as $\boldsymbol{\Omega}=\left[\boldsymbol{\omega}_1,\boldsymbol{\omega}_2,\cdots,\boldsymbol{\omega}_B\right]$. 

\subsection{Problem Formulation}
Given the defined system model, our goal is to simultaneously minimize the training loss on the multi-modal unlabeled dataset $\mathcal{D}^{\textrm{M}}$ and labeled dataset $\mathcal{D}_b^{\textrm{L}}$ at each BS. In particular, the training loss on $\mathcal{D}_b^{\textrm{L}}$ can be captured by the mean square error (MSE) \cite{36} between the predicted vehicle locations and labels, which is given by
\begin{equation}\label{eq:Pretraining Dataset}
L_b^{\textrm{L}} \left(\boldsymbol{\omega}_b\right)=\frac{1}{N_{b}^{\textrm{L}}}\sum_{j=1}^{N_{b}^{\textrm{L}}} \| \boldsymbol{p}_{b,j} - F_{\boldsymbol{\omega}_b}\left(\boldsymbol{H}_{b,j}\right)\|_2^2 ,
\end{equation}
where $ \| \cdot \|_2$ denotes the $L2$-norm of a vector. The training loss on the the multi-modal dataset $\mathcal{D}^{\textrm{M}}$ cannot be directly captured by MSE since we do not know which $\boldsymbol{p}_{j}\in\mathcal{P}^k$ is the true label of $\boldsymbol{H}_{b,v}\in\mathcal{H}^k$. To derive the loss function on $\mathcal{D}^{\textrm{M}}$, we first use $\boldsymbol{\alpha}_{b,v}=\left[\alpha_{b,v}^{1},\alpha_{b,v}^{2},\cdots,\alpha_{b,v}^{\hat{V}},\alpha_{b,v}^{\hat{V} +1}\right]^{\textrm{T}}$ with $\alpha_{b,v}^{j}\in\left\{0,1\right\}$, to represent the corresponding relationship between each $\boldsymbol{H}_{b,v}$ and $\boldsymbol{p}_{j}$, where $\alpha_{b,v}^{j}=1 \left(j=1,2,\cdots,\hat{V}\right)$ implies that the coordinate $\boldsymbol{p}_{j}$ is the corresponding positioning label of $\boldsymbol{H}_{b,v}$, and otherwise, $\alpha_{b,v}^{j}=0$, $\alpha_{b,v}^{\hat{V} +1}=1$ indicates that the corresponding label of $\boldsymbol{H}_{b,v}$ is not in $\mathcal{P}^k$. Given $\boldsymbol{\alpha}_{b,v}$, the corresponding location label of $\boldsymbol{H}_{b,v}$ is 
\begin{equation}
\boldsymbol{r}_{b,v}\left(\boldsymbol{\alpha}_{b,v}\right)=\alpha_{b,v}^{\hat{V} +1}F_{\boldsymbol{\omega}_b}\left(\boldsymbol{H}_{b,v}\right) + 
\left(1-\alpha_{b,v}^{\hat{V} +1}\right)\sum_{j=1}^{\hat{V}} \alpha_{b,v}^{j} \boldsymbol{p}_{j}.
\end{equation} 
From (4), we see that if $\boldsymbol{H}_{b,v}$ has a corresponding location in $\mathcal{P}^{k}$, then $\boldsymbol{r}_{b,v}\left(\boldsymbol{\alpha}_{b,v}\right)$ is equal to its corresponding $\boldsymbol{p}_{j}\in\mathcal{P}^{k}$. If $\mathcal{P}^{k}$ does not include the corresponding location label of $\boldsymbol{H}_{b,v}$, we directly use the vehicle location $F_{\boldsymbol{\omega}_b}\left(\boldsymbol{H}_{b,v}\right)$ predicted by NN as the positioning label of $\boldsymbol{H}_{b,v}$. 
We assume that the set of $\boldsymbol{\alpha}_{b,v}$ of all $\boldsymbol{H}_{b,v}\in\mathcal{H}^k$ is $\mathcal{A}^k$, and hence, we have $\boldsymbol{\alpha}_{b,v}\in\mathcal{A}^k$ for all possible $b,v$. Therefore, the loss function on $\left\{\mathcal{H}^k , \mathcal{P}^k\right\}\in\mathcal{D}^{\textrm{M}}$ at BS $b$ can be expressed as the MSE between the corresponding location label $\boldsymbol{r}_{b,v}\left(\boldsymbol{\alpha}_{b,v}\right)$ of each unlabeled $\boldsymbol{H}_{b,v}$ and its location $F_{\boldsymbol{\omega}_b}\left(\boldsymbol{H}_{b,v}\right)$ predicted by NN, which is given by
\begin{equation}
L_b^{\textrm{M}} \left(\boldsymbol{\omega}_b , \mathcal{A}^k\right) =\frac{\sum_{v=1}^{V_b} 
\| \boldsymbol{r}_{b,v}\left(\boldsymbol{\alpha}_{b,v}\right) - F_{\boldsymbol{\omega}_b}\left(\boldsymbol{H}_{b,v}\right)\|_2^2}{\sum_{v=1}^{V_b} \left(1 - \alpha_{b,v}^{\hat{V} +1}\right)}, 
\end{equation}
where $\sum_{v=1}^{V_b} \left(1 - \alpha_{b,v}^{\hat{V} +1}\right)$ is the number of unlabeled CSI in $\mathcal{H}^k$ which have their corresponding location labels in $\mathcal{P}^k$. Then, we formulate the training objective at BS $b$ as the linear combination of the MSE loss on $\mathcal{D}^{\textrm{M}}$ and $\mathcal{D}_b^{\textrm{L}}$, which is given by
\begin{equation}\label{target_problem}
\begin{aligned}
\boldsymbol{\omega}_b^{*}= \mathop{\arg\min}\limits_{\boldsymbol{\omega}_b} \, \Big\{ \gamma\cdot 
\Big[\frac{1}{N^{\textrm{M}}} \sum_{k=1}^{N^{\textrm{M}}}
L_b^{\textrm{M}} \left(\boldsymbol{\omega}_b ,\mathcal{A}^k\right)\Big]
\\
+ \left(1-\gamma\right)\cdot L_b^{\textrm{L}} \left(\boldsymbol{\omega}_b\right)\Big\},
\end{aligned}
\end{equation}
where $\boldsymbol{\omega}_b^{*}$ denotes the optimal NN parameter of BS $b$, and $\gamma$ is a weight parameter.
From (6), we see that the optimal $\boldsymbol{\omega}_b$ jointly depends on $\gamma$ and $\mathcal{A}^k$, since assigning different labels to the same unlabeled CSI will result in different training models, and the weight parameter $\gamma$ can be used to adjust the confidence level towards the labels of CSI samples assigned by $\mathcal{A}^{k}$. For instance, a larger $\gamma$ should be selected if the assigned position labels are very close to the true position labels of the unlabeled CSI. Given (5) and (6), the training objective of all BSs is formulated as the following bi-level optimization problem \cite{Bilevel}:
\begin{equation}\label{optimization}
\begin{aligned}
\mathop{\min}\limits_{\boldsymbol{\Omega}} \, \sum_{b=1}^{B}\Big\{
\gamma\cdot 
\Big[\frac{1}{N^{\textrm{M}}} \sum_{k=1}^{N^{\textrm{M}}}
L_b^{\textrm{M}} \left(\boldsymbol{\omega}_b ,\hat{\mathcal{A}}^k\right)\Big]\;\;\;\;\;\;\;\;\;\;\;\;\;\;\;\;\;\;\;\;\;\;\;\;\;\;&
\\
+ \left(1-\gamma\right)\cdot L_b^{\textrm{L}} \left(\boldsymbol{\omega}_b\right)\Big\}, \;\;\;\;\;\;\textrm{(7)}&
\\
{\rm{s.t.}} \,\,\,\, \hat{\mathcal{A}}^k =\;\;\;\;\;\;\;\;\;\;\;\;\;\;\;\;\;\;\;\;\;\;\;\;\;\;\;\;\;\;\;\;\;\;\;\;\;\;\;\;\;\;\;\;\;\;\;\;\;\;\;\;\;\;&
\\
\,\;\;\;\;\;\;\;\;\mathop{\arg\min}\limits_{\mathcal{A}^k} \, \sum_{b=1}^{B}\sum_{v=1}^{V_b}  \frac{
\| \boldsymbol{r}_{b,v}\left(\boldsymbol{\alpha}_{b,v}\right) - F_{\boldsymbol{\omega}_b}\left(\boldsymbol{H}_{b,v}\right)\|_2^2}{\sigma_{\boldsymbol{\omega}_b}^{2}\left(\boldsymbol{H}_{b,v}\right)},  \;\;\;\;\;\; \textrm{(7a)}&
\\
\alpha_{b,v}^{j}\in\big\{0,1\big\}, \;\;\;\;\;\;\;\;\;\;\;\;\;\;\;\;\;\;\;\;\; \;\;\;\;\;\;\;\;\;\textrm{(7b)}&
\\  
\sum_{b=1}^{B} \sum_{v=1}^{V_b}  \alpha_{b,v}^{j} =  1,  \;\;\;\;\;\;\;\;\;\;\;\;\;\;\;\;\;\; \;\;\;\;\;\;\;\;
 \textrm{(7c)}&
\\
\sum_{j=1}^{\hat{V} +1} \alpha_{b,v}^{j} = 1, \;\;\;\;\;\;\;\;\;\;\;\;\;\;\;\;\;\;\;\;\;\;\;\;\;\;\;\;\;\;\;\textrm{(7d)}&
\end{aligned}
\nonumber
\end{equation}
where (7a) implies that the locations predicted by the trained models should be closest to the corresponding vehicle locations obtained from images, and $\sigma_{\boldsymbol{\omega}_{b}} \left(\boldsymbol{H}_{b,v}\right)$ in (7a) represents the positioning error at the predicted location of $\boldsymbol{H}_{b,v}$, which is used to alleviate the impact of low positioning accuracy regions \cite{35}. Specifically, if the positioning accuracy in a certain region is low, the probability of correctly matching the vehicles locations obtained from images to the CSI samples collected in this region is low. Hence, we decrease the weights of the MSE of these data samples by using $\sigma_{\boldsymbol{\omega}_{b}} \left(\boldsymbol{H}_{b,v}\right)$. (7b) indicates whether the corresponding vehicle location of CSI $\boldsymbol{H}_{b,v}$ is position coordinate $\boldsymbol{p}_{j}$, (7c) guarantees that each position coordinate can only correspond to one CSI, and (7d) ensures that each CSI has at most one corresponding location coordinate since the vehicle may not be captured by all cameras. From (7), we see that the the objective function jointly depends on the training loss on the labeled dataset $\mathcal{D}_b^{\textrm{L}}$ and the multi-modal dataset $\mathcal{D}^{\textrm{M}}$. Moreover, since the corresponding relationships between CSI and vehicle locations in images are unknown to us, the selection of $\mathcal{A}^k$ is further constrained by the optimal solution of the optimization problem in (7a). In consequence, the minimization of (7) cannot be directly achieved by gradient descent algorithms. To avoid the difficulties of straightforwardly solving the bi-level optimization problem in (7), we propose a meta-learning based hard expectation-maximization (EM) algorithm, which alternately finds the optimal corresponding relationship vectors in (7a) and optimizes the model parameters in (7).

\section{Proposed Meta-learning based Hard EM Algorithm}
In this section, we introduce the proposed meta-learning based hard EM algorithm that can effectively solve (7). Compared to the current positioning algorithms, the proposed algorithm can jointly use images and unlabeled CSI to train the positioning model, thus achieving higher positioning accuracy when the number of labeled training samples is small. Compared to standard EM algorithms, our algorithm retains the low computational complexity of hard EM, while additionally introducing a meta-learning based weighting approach to improve the convergence of model. Next, we first introduce the three main steps in each training iteration of the proposed algorithm: 1) construction of the training objective at each iteration, 2) optimization of the objective function, 3) meta-learning based weighting. Then, the inference procedure in testing stage is presented. Finally, we discuss the computational complexity and convergence of the proposed algorithm.

\subsection{Main Steps}

\subsubsection{Construction of the training objective}
The hard EM algorithm can be seen as a cyclic coordinate optimization method that alternately estimates $\hat{\mathcal{A}}^k$ in (7a) and optimizes the model parameter $\boldsymbol{\Omega}$ in (7). Specifically, we first find the locally optimal solution $\hat{\mathcal{A}}^k$ of (7a) when given the model parameters $\boldsymbol{\Omega}^{i-1}=\left[\boldsymbol{\omega}_{1}^{i-1},\cdots,\boldsymbol{\omega}_{B}^{i-1}\right]$ at iteration $i-1$. Then, the estimated $\hat{\mathcal{A}}^k$ are substituted into (7) to obtain a temporary training objective at iteration $i$. Next, we introduce the algorithm for solving (7a). In particular, given $\boldsymbol{\Omega}^{i-1}$ at iteration $i-1$, the problem in (7a) can be simplified as
\begin{equation}\label{optimization}
\begin{aligned}
\;\;\;\;\;\;\;\;\,\,\mathop{\min}\limits_{\mathcal{A}^k}  \, \sum_{b=1}^{B} &\sum_{v=1}^{V_b} \frac{\| \boldsymbol{r}_{b,v}\left(\boldsymbol{\alpha}_{b,v}\right) - F_{\boldsymbol{\omega}_b^{i-1}}\left(\boldsymbol{H}_{b,v}\right)\|_2^2}{
\hat{\sigma}_{\boldsymbol{\omega}_{b}^{i-1}}^{2} \left(\boldsymbol{H}_{b,v}\right)
}, \;\;\;\;\;\;\;\;\;\; \textrm{(8)}
\\
&{\rm{s.t.}} \,\,\,\,\textrm{(7b), (7c), and (7d)},
\end{aligned}
\nonumber
\end{equation}
where $\hat{\sigma}_{\boldsymbol{\omega}_{b}^{i-1}} \left(\boldsymbol{H}_{b,v}\right)$ denotes the empirical positioning error at the predicted location of $\boldsymbol{H}_{b,v}$. Since (8) is a standard minimum matching problem, we can use the Kuhn-Munkres (KM) algorithm \cite{39} to solve (8). To define the input of KM algorithm, we first define the cost of matching $\boldsymbol{H}_{b,v}\in\mathcal{H}^k$ to $\boldsymbol{p}_{j}\in\mathcal{P}^k$ as
\begin{equation}
R_{\boldsymbol{\omega}_{b}^{i-1}}\left(\boldsymbol{H}_{b,v},\boldsymbol{p}_{j}\right)= \frac{\| F_{\boldsymbol{\omega}_b^{i-1}}\left(\boldsymbol{H}_{b,v}\right) -  \boldsymbol{p}_{j}  \|_2^2}{
\hat{\sigma}_{\boldsymbol{\omega}_{b}^{i-1}}^{2} \left(\boldsymbol{H}_{b,v}\right)
}. \tag{9}
\end{equation}
Given (9), the input of KM algorithm is defined as a cost matrix $\boldsymbol{C}_{i}\in\mathbb{R}^{V \times \hat{V}}$ where each element $c_{i,m,n}$ at row $m$ and column $n$ represents the cost of matching the CSI $m$ in $\mathcal{H}^k$ to the location coordinate $n$ in $\mathcal{P}^k$. Here, $V = \sum_{b=1}^B V_b$ is the number of CSI matrices in $\mathcal{H}^k$. Hence, we have $c_{i,m,n}  = R_{\boldsymbol{\omega}_{b_m}^{i-1}}\left(\boldsymbol{H}_{b_m,v_m},\boldsymbol{p}_{n}\right)$. Given $\boldsymbol{C}_{i}$, the optimal corresponding relationship vector $\hat{\boldsymbol{\alpha}}_{b , v}=\left[\hat{\alpha}_{b,v}^{1},\hat{\alpha}_{b,v}^{2},\cdots,\hat{\alpha}_{b,v}^{\hat{V}},\hat{\alpha}_{b,v}^{\hat{V} +1}\right]^{\textrm{T}}$ of each $\boldsymbol{H}_{b , v}$ at iteration $i$ can be obtained by using the KM algorithm. Given $\hat{\boldsymbol{\alpha}}_{b , v}$, the set of corresponding location labels of the unlabeled CSI samples at iteration $i$ is
\begin{equation}
\begin{aligned}
\hat{\mathcal{R}}_{\boldsymbol{\Omega}^{i-1}}^{k}=\Big\{\hat{\boldsymbol{r}}_{b,v}\left(\hat{\boldsymbol{\alpha}}_{b,v}\right) \Big|
\hat{\boldsymbol{r}}_{b,v}\left(\hat{\boldsymbol{\alpha}}_{b,v}\right) = \hat{\alpha}_{b,v}^{\hat{V} +1}\hat{\boldsymbol{p}}_{b,v}
\\
+ 
\left(1-\hat{\alpha}_{b,v}^{\hat{V} +1}\right)\sum_{j=1}^{\hat{V}} \hat{\alpha}_{b,v}^{j} \boldsymbol{p}_{j} \Big\},
\end{aligned}\tag{10}
\end{equation}
where $\hat{\boldsymbol{r}}_{b,v}\left(\hat{\boldsymbol{\alpha}}_{b,v}\right)$ denotes the position label of $\boldsymbol{H}_{b,v}\in\mathcal{H}^k$. Substituting (10) into (7) we can obtain the training objective at iteration $i$, which is expressed as
\begin{equation}\label{lowerBoundMaximization}
\begin{aligned}
\mathop{\min}\limits_{\boldsymbol{\Omega}} \sum_{b=1}^{B}\bigg\{
\gamma\cdot 
\Big[\frac{1}{{N^{\textrm{M}}}}  \sum_{k=1}^{N^{\textrm{M}}} & \frac{\sum_{v=1}^{V_b} 
\| \hat{\boldsymbol{r}}_{b,v}\left(\hat{\boldsymbol{\alpha}}_{b,v}\right) - F_{\boldsymbol{\omega}_b}\left(\boldsymbol{H}_{b,v}\right)\|_2^2}{
\sum_{v=1}^{V_b} \left(1 - \hat{\alpha}_{b,v}^{\hat{V} +1}\right)}\Big]
\\
&+ \left(1-\gamma\right)\cdot L_b^{\textrm{L}} \left(\boldsymbol{\omega}_b\right)\bigg\}.
\end{aligned}\tag{11}
\end{equation}


\subsubsection{Optimization of the objective function}
Since the optimization problem in (11) does not depend on constraints (7a) - (7d), the problem in (11) is an unconstrained problem, and hence, (11) can be directly solved by utilizing a gradient descent algorithm \cite{40,41}. Therefore, the updated NN parameter at BS $b$ can be expressed as
\begin{equation}\label{eq:MSE Loss 1}
\begin{aligned}
&\boldsymbol{\omega}_{b}^{i}  \gets\boldsymbol{\omega}_{b}^{i-1}-
\\
&\lambda_{i} \nabla_{\boldsymbol{\omega}} \bigg\{
\gamma
\bigg[\frac{1}{{N^{\textrm{M}}}}  \sum_{k=1}^{N^{\textrm{M}}} \frac{\sum_{v=1}^{V_b} 
\| \hat{\boldsymbol{r}}_{b,v}\left(\hat{\boldsymbol{\alpha}}_{b,v}\right) 
- F_{\boldsymbol{\omega}}\big(\boldsymbol{H}_{b,v}\big)\|_2^2}{
\sum_{v=1}^{V_b} \big(1 - \hat{\alpha}_{b,v}^{\hat{V} +1}\big)}\bigg]
\\
&  \;\;\;\;\;\;\;\;   + \left(1-\gamma\right) L_b^{\textrm{L}} \left(\boldsymbol{\omega}\right)\bigg\} \bigg|_{\boldsymbol{\omega}=\boldsymbol{\omega}_{b}^{i-1}}\;\; , 
\end{aligned}\tag{12}
\end{equation}
where $\lambda_i$ is the learning rate at iteration $i$, and $\nabla$ denotes the gradient operator.


\subsubsection{Meta-learning based weighting}
To reduce the impact of label noises caused by incorrect matching between unlabeled CSI and vehicle locations obtained from images so as to achieve better convergence, we introduce a weight for the loss of each data sample (i.e. $\| \hat{\boldsymbol{r}}_{b,v}\left(\hat{\boldsymbol{\alpha}}_{b,v}\right) - F_{\boldsymbol{\omega}}\left(\boldsymbol{H}_{b,v}\right)\|_2^2$) in $\mathcal{D}^{\textrm{M}}$, and study the use of a meta-learning algorithm \cite{38} for computing this weighted loss on $\mathcal{D}^{\textrm{M}}$. Specifically, the weighted loss on $\mathcal{D}^{\textrm{M}}$ is defined as
\begin{equation}
\frac{1}{{N^{\textrm{M}}}}  \sum_{k=1}^{N^{\textrm{M}}} \frac{\sum_{v=1}^{V_b} 
\hat{\epsilon}_{b,v}\| \hat{\boldsymbol{r}}_{b,v}\left(\hat{\boldsymbol{\alpha}}_{b,v}\right) - F_{\boldsymbol{\omega}_b}\left(\boldsymbol{H}_{b,v}\right)\|_2^2}{
\sum_{v=1}^{V_b} \left(1 - \hat{\alpha}_{b,v}^{\hat{V} +1}\right)}, \tag{13}
\end{equation}
where $\hat{\epsilon}_{b,v}$ denotes the weight parameter of the CSI sample $\boldsymbol{H}_{b,v}\in\mathcal{H}^k$. Intuitively, we expect to reduce the effect of label noises by assigning a smaller $\hat{\epsilon}_{b,v}$ to the case where $\hat{\boldsymbol{r}}_{b,v}\left(\hat{\boldsymbol{\alpha}}_{b,v}\right)$ is not the true label of $\boldsymbol{H}_{b,v}$. Next, we introduce a meta-learning algorithm for computing each weight parameter. Specifically, we first define the NN parameter updated on $\left\{\mathcal{H}^k,\hat{\mathcal{R}}_{\boldsymbol{\Omega}^{i-1}}^{k}\right\}$ with gradient descent method as a function of the weight parameters vector $\boldsymbol{\epsilon}_b = \left[\epsilon_{b,1},\epsilon_{b,2},\cdots,\epsilon_{b,V_b}\right]^{\textrm{T}}\in\mathbb{R}^{V_b \times 1}$, which is given by
\begin{equation}
\begin{aligned}
&\hat{\boldsymbol{\omega}}_{b}^{i}\left(\boldsymbol{\epsilon}_b\right)=\boldsymbol{\omega}_{b}^{i-1}
-
\\
&\lambda_{i} \nabla_{\boldsymbol{\omega}}\Big(\sum_{v=1}^{V_b} 
\epsilon_{b,v}\| \hat{\boldsymbol{r}}_{b,v}\left(\hat{\boldsymbol{\alpha}}_{b,v}\right) - F_{\boldsymbol{\omega}}\left(\boldsymbol{H}_{b,v}\right)\|_2^2   \Big)\Big|_{\boldsymbol{\omega}=\boldsymbol{\omega}_{b}^{i-1}}.
\end{aligned}\tag{14}
\end{equation}
To obtain the meta-learning objective, we then define a validation dataset $\mathcal{D}_{b}^{\textrm{V}}=\left\{ \boldsymbol{H}_{b,j}^{\textrm{V}},\boldsymbol{p}_{b,j}^{\textrm{V}}\right\}_{j=1}^{N_b^{\textrm{V}}}$ which is separated from the labeled dataset $\mathcal{D}_{b}^{\textrm{L}}$ at BS $b$, with $N_b^{\textrm{V}}$ being the number of validation samples. Given $\hat{\boldsymbol{\omega}}_{b}^{i}\left(\boldsymbol{\epsilon}_b\right)$ and $\mathcal{D}_{b}^{\textrm{V}}$, the meta-learning objective of determining $\hat{\boldsymbol{\epsilon}}_b$ is formulated as
\begin{equation}\label{optimization}
\begin{aligned}
 \;\;\;\;\;\;\;\;\;\;\;\;\;\;\mathop{\min}\limits_{\boldsymbol{\epsilon}_b}  \frac{1}{N_b^{\textrm{V}}}\sum_{j=1}^{N_{b}^{\textrm{V}}}
\| \boldsymbol{p}_{b,j}^{\textrm{V}} - F_{\hat{\boldsymbol{\omega}}_{b}^{i}\left(\boldsymbol{\epsilon}_b\right)}\left(\boldsymbol{H}_{b,j}^{\textrm{V}}\right)\|_2^2  , \;\;\;\;\;\;\;\;\;\;\; \textrm{(15)}&
\\
{\rm{s.t.}} \, \epsilon_{b,v} \geq 0, \;\;\;\;\;\;\;\;\;\;\;\;\;\;\;\;\;\;\;\;\;\;\;\;\; \textrm{(15a)}&
\end{aligned}
\nonumber
\end{equation}
where the optimization objective in (15) indicates that the optimum $\boldsymbol{\epsilon}_b$ should be able to achieve the lowest positioning error, and (15a) guarantees that all the weighting coefficients are non-negative. Considering that directly solving (15) is time-consuming, we only take a single gradient descent step on the validation dataset with regard to $\boldsymbol{\epsilon}_b$ in order to obtain an approximation of the optimal weighting coefficients at iteration $i$, which is given by
\begin{equation}
\begin{aligned}
\widetilde{\epsilon}_{b,v} &=-\frac{\partial}{\partial \epsilon_{b,v}} \Big(\frac{\sum_{j=1}^{N_{b}^{\textrm{V}}}
\| \boldsymbol{p}_{b,j}^{\textrm{V}} - F_{\hat{\boldsymbol{\omega}}_{b}^{i}\left(\boldsymbol{\epsilon}_b\right)}\left(\boldsymbol{H}_{b,j}^{\textrm{V}}\right)\|_2^2}{N_b^{\textrm{V}}} \Big)\Big|_{\epsilon_{b,v}=0}\\
&\;\;\;\;\;=\bigg\langle    \nabla_{\boldsymbol{\omega}}\frac{\sum_{j=1}^{N_{b}^{\textrm{V}}}
\| \boldsymbol{p}_{b,j}^{\textrm{V}} - F_{\boldsymbol{\omega}}\left(\boldsymbol{H}_{b,j}^{\textrm{V}}\right)\|_2^2}{N_{b}^{\textrm{V}}},
\\
& \;\;\;\;\;\;\;\;\;\;\;\;\;\;\; \nabla_{\boldsymbol{\omega}}
\| \hat{\boldsymbol{r}}_{b,v}\left(\hat{\boldsymbol{\alpha}}_{b,v}\right) - F_{\boldsymbol{\omega}}\left(\boldsymbol{H}_{b,v}\right)\|_2^2
\bigg\rangle \bigg|_{\boldsymbol{\omega}=\boldsymbol{\omega}_{b}^{i-1}}\;\; , 
\end{aligned} \tag{16}
\end{equation}
with $\left\langle \cdot\, ,\cdot \right\rangle$ being the inner product of two vectors. Since $\widetilde{\epsilon}_{b,v}$ may be a negative number, we may need to adjust the value of $\widetilde{\epsilon}_{b,v}$ to satisfy the constrain (15a) when $\widetilde{\epsilon}_{b,v}$ is negative. Here, we propose a new method defined as
\begin{equation}
\hat{\epsilon}_{b,v}=1 + \xi\cdot\frac{\widetilde{\epsilon}_{b,v}}{\mathop{\max}\limits_{v}
\vert \widetilde{\epsilon}_{b,v} \vert },\tag{17}
\end{equation}
where $\xi\in\left[0,1\right]$ is a weight parameter. The proposed method (17) can achieve better performance in avoiding overfitting compared to the rectified linear unit (ReLU) method \cite{38}, as shown in our simulation results in Section $\textrm{\uppercase\expandafter{\romannumeral4}}$. Given the weighted MSE loss, we use the gradient descent method to optimize the NN parameter at BS $b$.

The specific meta-learning based hard EM algorithm is summarized in \textbf{Algorithm~1}.

\begin{algorithm}[t]\setstretch{1.2}
\footnotesize 
\caption{Meta-learning based hard EM algorithm.}
\begin{algorithmic}[1]
\STATE \textbf{Initialize:} Model parameters $\boldsymbol{\Omega}^0$, learning rate $\lambda$, rectification scale $\xi$, weight parameter $\gamma$,  the number of training iteration $T$, iteration period $N$ of re-estimating $\sigma_{\omega_{b}} \left(\cdot\right)$.

\STATE \textbf{Input:} Training datasets $\mathcal{D}_{b}^{\textrm{L}}$, validation datasets $\mathcal{D}_{b}^{\textrm{V}}$, multi-modal dataset $\mathcal{D}^{\textrm{M}}$
\FOR {$i = 1 \to T$} 
\IF{$i-1 \equiv 0 \pmod{N}$}
\STATE Estimate $\hat{\sigma}_{\omega_{b}} \left(\cdot\right)$ on validation datasets $\mathcal{D}_{b}^{\textrm{V}}$.
\ENDIF

\STATE Calculate $\boldsymbol{C}_{i}$ based on (9).
\STATE Solve each $\hat{\boldsymbol{\alpha}}_{b,v}$ with KM algorithm.
\STATE Obtain $\hat{\mathcal{R}}_{\boldsymbol{\Omega}^{i}}^{k}$ based on (10).
\STATE Calculate each $\widetilde{\epsilon}_{b,v}$ based on (16).
\STATE Adjust the value of $\widetilde{\epsilon}_{b,v}$ to obtain $\hat{\epsilon}_{b,v}$ according to (17).
\STATE Calculate the weighted loss based on (13).
\STATE Update $\boldsymbol{\omega}_{b}$ by (12).
\ENDFOR
\end{algorithmic}
\label{algorithm_1}
\end{algorithm}

\subsection{Inference Procedure} 
In the testing stage, the locations of vehicles are jointly determined by CSI and images. In particular, at each time slot, BS $b$ first utilizes its NN to predict the position coordinate of each $\boldsymbol{H}_{b,v}$, and obtains the position coordinates of the vehicles from its captured images. Then, BS $b$ will share all the predicted positions of CSI and vehicle locations obtained from images with other BSs, such that each BS can solve problem (8) in a distributed manner. Given $\hat{\boldsymbol{\alpha}}_{b,v}$ obtained by solving (8), the position coordinate of $\boldsymbol{H}_{b,v}$ is
\begin{equation}
\label{penalty}
\widetilde{\boldsymbol{p}}_{b,v} = \left\{ {\begin{array}{*{20}{l}}
{F_{\boldsymbol{\omega}_b}\left(\boldsymbol{H}_{b,v}\right),\;\;\;\;\;\;\;\hat{\alpha}_{b,v}^{\hat{V} +1}=1,}\\
{\sum_{j=1}^{\hat{V}} \hat{\alpha}_{b,v}^{j} \boldsymbol{p}_{j}, \;\;\;\;\;\hat{\alpha}_{b,v}^{\hat{V} +1}=0.}
\end{array}} \right. \tag{18}
\end{equation}

\subsection{Complexity and convergence Analysis} 
In this section, we analyze the computational complexity and convergence of the proposed algorithm. Here, considering that the complexity of the designed algorithm depends on the structure of the NN models, we first make some assumptions on the NN structure. Specifically, we select to use CNN as the positioning model at each BS. We assume that the CNN model at each BS consists of $M^{\textrm{C}}$ convolutional layers and $M^{\textrm{F}}$ fully connected layers. For convolutional layer $m$, we assume that the input dimension is $c_{m}^{\textrm{I}} \times w_{m}^{\textrm{I}} \times h_{m}^{\textrm{I}}$, with $c_{m}^{\textrm{I}}$ being the number of input channels, $w_{m}^{\textrm{I}}$ and $h_{m}^{\textrm{I}}$ respectively being the width and height of the input data. Correspondingly, the output dimension of convolutional layer $m$ is $c_{m}^{\textrm{O}} \times w_{m}^{\textrm{O}} \times h_{m}^{\textrm{O}}$, where $c_{m}^{\textrm{O}}$, $w_{m}^{\textrm{O}}$, and $h_{m}^{\textrm{O}}$ respectively denote the number of output channels, the width and height of the output data. The spatial size of the convolutional kernel of convolutional layer $m$ is represented by $s_{m}$. For fully connected layer $n$, we assume that the input dimension is $f_{n}^{\textrm{I}}$, and the output dimension is $f_{n}^{\textrm{O}}$.

Next, we introduce the complexity of the proposed algorithm which mainly lies in: 1) KM algorithm for solving (8), 2) meta-learning algorithm for determining $\hat{\boldsymbol{\epsilon}}_{b}$, and 3) training the CNN-based positioning model. Given $\left\{\mathcal{H}^k,\mathcal{P}^k\right\}\in\mathcal{D}^{\textrm{M}}$, the computational complexity of solving (8) with KM algorithm is $\mathcal{O}\left(V^3\right)$ according to \cite{KuhnMunkres}, where $V$ is the number of CSI samples in $\mathcal{H}^k$. The complexity of meta-learning mainly depends on calculating the gradient vectors on validation and training samples. Hence, according to \cite{38} and \cite{37}, the computational complexity of meta-learning is $\mathcal{O}\left(\left(N_b^{\textrm{V}} + V\right)\left(\sum_{m} c_{m}^{\textrm{I}} c_{m}^{\textrm{O}} w_{m}^{\textrm{I}} h_{m}^{\textrm{I}} s_{m}^2 + \sum_{n} f_{n}^{\textrm{I}} f_{n}^{\textrm{O}}  \right)\right)$. Given the weight parameters solved by using the meta-learning algorithm, the complexity for training the CNN-based positioning model is given by $\mathcal{O}\left(V\left(\sum_{m} c_{m}^{\textrm{I}} c_{m}^{\textrm{O}} w_{m}^{\textrm{I}} h_{m}^{\textrm{I}} s_{m}^2 + \sum_{n} f_{n}^{\textrm{I}} f_{n}^{\textrm{O}}  \right)\right)$ \cite{37}. Since we always have $V^2 \ll \sum_{m} c_{m}^{\textrm{I}} c_{m}^{\textrm{O}} w_{m}^{\textrm{I}} h_{m}^{\textrm{I}} s_{m}^2 + \sum_{n} f_{n}^{\textrm{I}} f_{n}^{\textrm{O}}$ in practical scenarios, the complexity of the proposed algorithm is dominated by the meta-learning algorithm and the process of training the CNN. Hence, the computational complexity of the proposed algorithm at each iteration is
\begin{equation}
\mathcal{O}\left(\left(N_b^{\textrm{V}} + 2V\right)\left(\sum_{m=1}^{M^{\textrm{C}}} c_{m}^{\textrm{I}} c_{m}^{\textrm{O}} w_{m}^{\textrm{I}} h_{m}^{\textrm{I}} s_{m}^2 + \sum_{n=1}^{M^{\textrm{F}}} f_{n}^{\textrm{I}} f_{n}^{\textrm{O}}  \right)\right).\tag{19}
\end{equation}

\begin{table}\footnotesize
\newcommand{\tabincell}[2]{\begin{tabular}{@{}#1@{}}#1.1\end{tabular}}
\renewcommand\arraystretch{1.3}
\caption[table]{{System Parameters}}
\centering
\begin{tabular}{|c|c|c|c|c|c|}
\hline
\!\textbf{Parameter}\! \!\!& \textbf{value} &\! \textbf{Parameter} \!& \textbf{Value} \\
\hline
$B$ & $2$ &  $C$ & $3$\\
\hline
$N^{\textrm{B}}$  & 16 &  $N^{\textrm{C}}$ & 52\\
\hline
$W$ & 1280 & $H$ & 720 \\
\hline
$\lambda$ & $10^{-3}$ & $\gamma$ & $0.5$ \\
\hline
$\xi$ & $1$ & $N_{\textrm{M}}$ & 2000 \\
\hline
$N_b^{\textrm{L}}$ & $300$ to $5000$ & $N_b^{\textrm{V}}$ & $\frac{1}{3} N_b^{\textrm{L}}$ \\
\hline
\end{tabular}
\label{table2}
\end{table}

Next, we discuss the convergence of the proposed algorithm. Following \cite{38}, we show that our method can converge to the critical point of the MSE loss on the validation dataset $\mathcal{D}_b^{\textrm{V}}$. We assume that $G_b \left(\boldsymbol{\omega}\right)=\frac{1}{N_{b}^{\textrm{V}}}\sum_{j=1}^{N_{b}^{\textrm{V}}}
\| \boldsymbol{p}_{b,j}^{\textrm{V}} - F_{\boldsymbol{\omega}}\left(\boldsymbol{H}_{b,j}^{\textrm{V}}\right)\|_2^2$, which represents the validation loss at BS $b$, $g_{b,v} \left(\boldsymbol{\omega}\right)=\| \hat{\boldsymbol{r}}_{b,v}\left(\hat{\boldsymbol{\alpha}}_{b,v}\right) - F_{\boldsymbol{\omega}}\left(\boldsymbol{H}_{b,v}\right)\|_2^2$ is the loss of unlabeled sample $\boldsymbol{H}_{b,v}$, $\mu_b=\mathop{\max}\limits_{v}\vert \widetilde{\epsilon}_{b,v} \vert$ is the maximum absolute value of $\widetilde{\epsilon}_{b,v}$, and $n_b=\sum_{v=1}^{V_b} \big(1 - \hat{\alpha}_{b,v}^{\hat{V} +1}\big)$ denotes the number of unlabeled CSI which have their corresponding location labels in $\mathcal{P}^k$. For simplicity, we prove the convergence in the situation where $\xi =1$ and $\gamma =1$, which means that we will ignore the influence of the labeled dataset $\mathcal{D}_b^{\textrm{L}}$ in our proof, and only consider the unlabeled multi-modal dataset $\mathcal{D}^{\textrm{M}}$ for model training in each iteration. We also assume that the NN parameter is updated on a batch of training data $\left\{\mathcal{H}^k,\hat{\mathcal{R}}_{\boldsymbol{\Omega}^{i-1}}^{k}\right\}$ with gradient descent method. To prove the convergence of our proposed algorithm, we first make the following assumptions, as done in \cite{38,42}.
\begin{itemize}
\item First, we assume that $G_b \left(\boldsymbol{\omega}_b\right)$ is uniformly Lipschitz continuous with respect to $\boldsymbol{\omega}_b$ \cite{38}. Hence, we have
\begin{equation}
\begin{aligned}
G_b \left(\boldsymbol{\omega}_b^{i+1}\right) \leq G_b \left(\boldsymbol{\omega}_b^{i}\right)&+
\left[\nabla G_b \left(\boldsymbol{\omega}_b^{i}\right)\right]^{\textrm{T}}\left(\boldsymbol{\omega}_b^{i+1} - \boldsymbol{\omega}_b^{i}\right)\\
&+\frac{L}{2}\| \boldsymbol{\omega}_b^{i+1} - \boldsymbol{\omega}_b^{i}\|_2^2,
\end{aligned}\tag{20}
\end{equation}
where $L$ is a positive constant.
\item Second, we assume that the gradients of both $G_b \left(\boldsymbol{\omega}\right)$ and $g_{b,v} \left(\boldsymbol{\omega}\right)$ are bounded by $\beta$, i.e. $\| \nabla G_b \left(\boldsymbol{\omega}\right)\|_2 \leq \beta$, and $\| \nabla g_{b,v} \left(\boldsymbol{\omega}\right)\|_2 \leq \beta$.
\item Since $\hat{\boldsymbol{r}}_{b,v}\left(\hat{\boldsymbol{\alpha}}_{b,v}\right)$ may be a wrong label for $\boldsymbol{H}_{b,v}$, there is a certain probability that the inner product of $\nabla_{\boldsymbol{\omega}}\frac{\sum_{j=1}^{N_{b}^{\textrm{V}}}
\| \boldsymbol{p}_{b,j}^{\textrm{V}} - F_{\boldsymbol{\omega}}\left(\boldsymbol{H}_{b,j}^{\textrm{V}}\right)\|_2^2}{N_{b}^{\textrm{V}}}$ and $\nabla_{\boldsymbol{\omega}}
\| \hat{\boldsymbol{r}}_{b,v}\left(\hat{\boldsymbol{\alpha}}_{b,v}\right) - F_{\boldsymbol{\omega}}\left(\boldsymbol{H}_{b,v}\right)\|_2^2$ (i.e. the weight parameter $\widetilde{\epsilon}_{b,v}$) can be negative-valued. For simplicity of analysis, we assume that except for $\widetilde{\epsilon}_{b,v}$ with the largest absolute value, the other $\widetilde{\epsilon}_{b,v}$ follow a uniform distribution from $-\mu_b$ to $\mu_b$, namely $\widetilde{\epsilon}_{b,v} \sim U\left(-\mu_b,\mu_b\right)$. We also assume that $V_b \gg 1$, i.e. BS $b$ has a large number of CSI samples in each $\mathcal{H}^k$.
\end{itemize}
Given the above assumptions, the convergence of the proposed algorithm can be obtained by the following lemma.
\itshape \textbf{Lemma 1:}  \upshape
If the learning rate $\lambda_i$ satisfies $\lambda_i\leq \frac{\mu_b n_b }{2L \beta^2}$, the validation loss $G_b \left(\boldsymbol{\omega}\right)$ remains monotonically non-increasing, namely
\begin{equation}
G_b \left(\boldsymbol{\omega}_b^{i+1}\right) \leq G_b \left(\boldsymbol{\omega}_b^{i}\right).\tag{21}
\end{equation}
Moreover, $\mathbb{E}\left[G_b \left(\boldsymbol{\omega}_b^{i+1}\right)-G_b \left(\boldsymbol{\omega}_b^{i}\right)\right]=0$ if and only if $\nabla G_b \left(\boldsymbol{\omega}_b^{i}\right)=0$, where the expectation is taking over all possible training samples in $\mathcal{D}^{\textrm{M}}$ at iteration $i$.

\itshape \text{Proof:}  \upshape
See Appendix A

From Lemma 1, we see that by controlling the learning rate $\lambda_i$ to be smaller than $\frac{\mu_b n_b }{2L \beta^2}$, the validation loss $G_b \left(\boldsymbol{\omega}\right)$ is monotonically non-increasing, and can converge to a stationary point.

\begin{figure}[t]
\centering
\setlength{\belowcaptionskip}{-0.5cm}
\includegraphics[width=8.5cm]{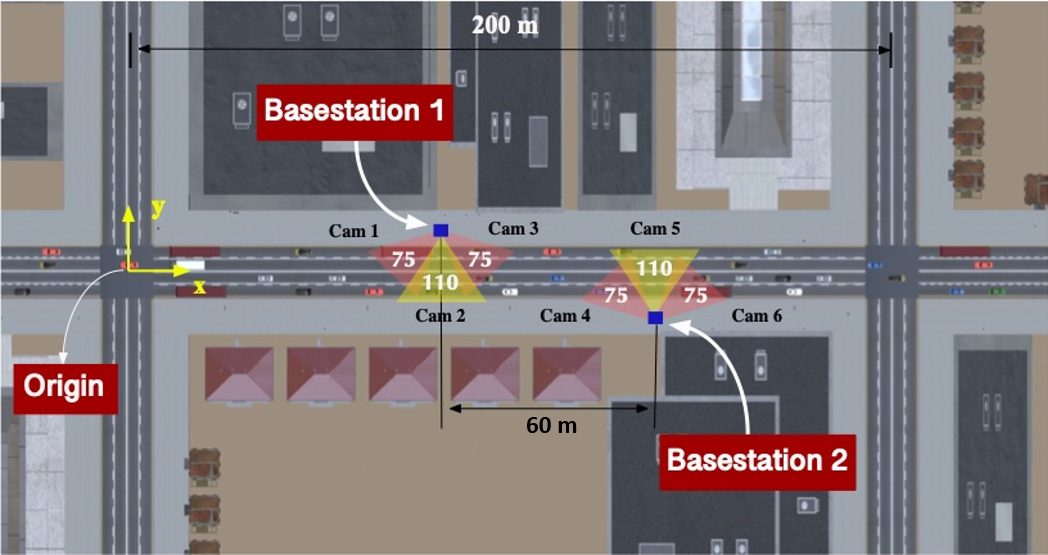}
\centering
\vspace{0.1cm}
\caption{Scenario of the dataset \cite{15}.}
\label{fig4}
\vspace{0.1cm}
\end{figure}


\section{Simulation results and analysis}
In this section, we evaluate the performance of our proposed positioning algorithm on a public dataset called Vision-Wireless (ViWi) \cite{15}. We first introduce the ViWi dataset used for model training. Then, we specify the experimental setup. Finally, we analyze the performance of our proposed algorithm.

\subsection{Dataset}
ViWi is a data-generating framework that not only provides wireless data of vehicles but combines it with visual data taken from the same scenes by utilizing advanced 3D modeling and ray-tracing simulators\cite{15}. As shown in Fig. \ref{fig4}, the dataset is generated in a downtown scenario with multiple served vehicles and two BSs situated at each side of the street. Each BS is equipped with three differently-oriented cameras such that the FoV can cover the whole street. Hence, each data sample in the dataset consists of six images from the two BSs, the CSI matrices, and position coordinates of the served vehicles. Here, considering that all the CSI matrices in $\mathcal{D}_{b}^{\textrm{L}}$ and $\mathcal{D}^{\textrm{M}}$ are complex-valued, we utilize the method in \cite{3} to transform them to real-valued matrices such that they can be used as inputs of the NNs. Since the vertical viewing angles of cameras are not given in the used dataset, we cannot obtain the elevation angle $\theta$ of a detected vehicle in image, and thus cannot calculate $d_{\textrm{H}}$. Therefore, we add zero-mean Gaussian noises with standard deviation of 1 m to the ground truth distances between vehicles and BSs, and use these disturbed distances as the vehicle-BS distances calculated from images. We assume that the distances on X-axis between the BS and its served vehicles are not larger than 55 m, and hence we filter out the locations obtained from images of the vehicles whose X-axis distances to their linked BSs are larger than 55 m. For the vehicles located in the overlapped serving area of the two BSs, we use the A3 event \cite{16} to determine their connected BSs. After down-sampling and discarding the unusable samples, we randomly select 2000 samples as $\mathcal{D}^{\textrm{M}}$, and 500 samples as the testing data. 


\subsection{Experimental Setup}
\subsubsection{Structure of NNs}
\begin{figure}[t]
\centering
\setlength{\belowcaptionskip}{-0.5cm}
\includegraphics[width=8.5cm]{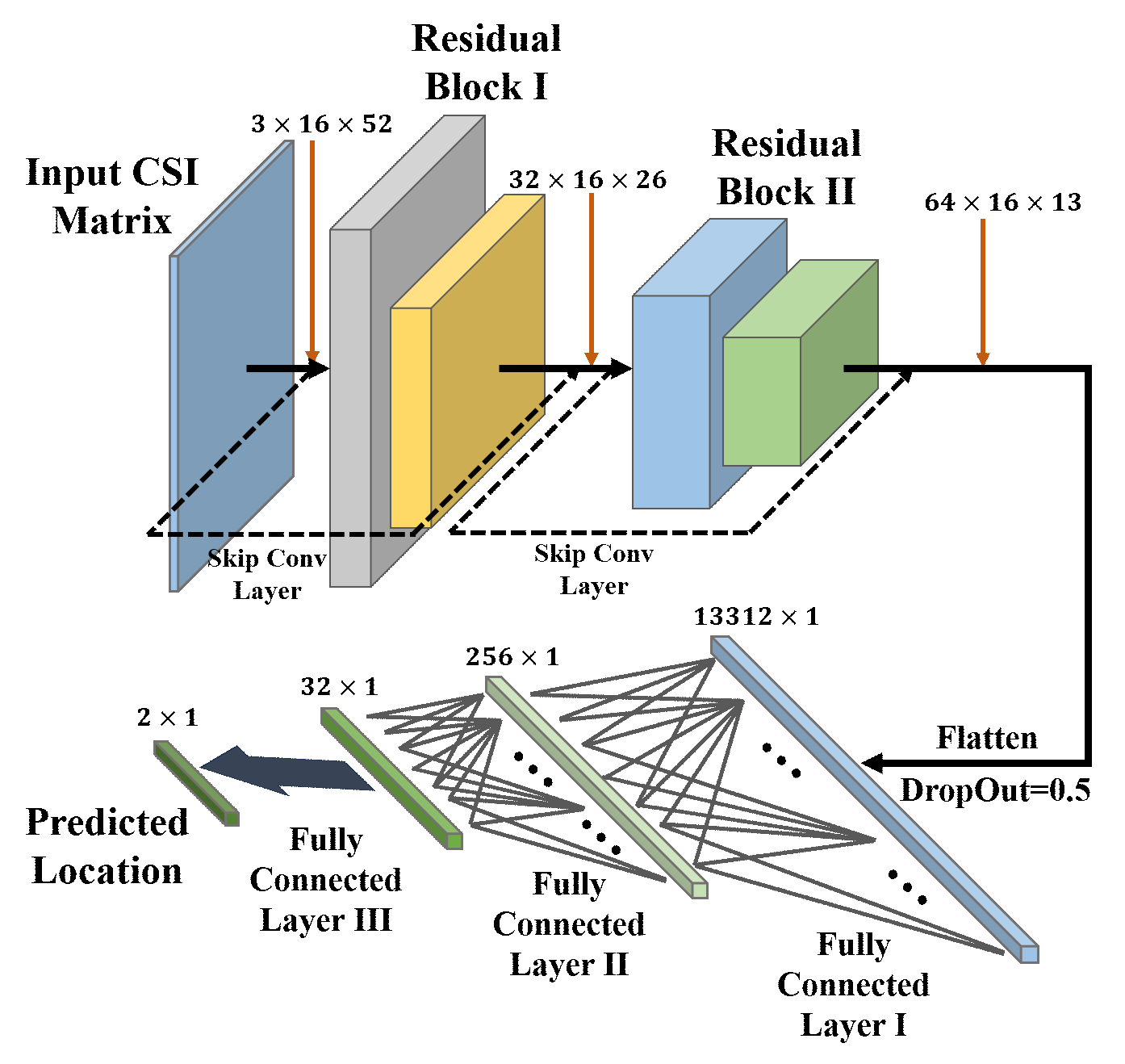}
\centering
\vspace{-0.1cm}
\caption{Structure of NNs.}
\label{fig5}
\vspace{-0.1cm}
\end{figure}

We assume that the NN at each BS has the same structure. As shown in Fig. \ref{fig5}, the employed NN model consists of two CNN-based residual blocks (RBs) \cite{14} and a fully connected network (FNN), which has been proven to be effective for RF fingerprint based positioning in \cite{4,14}. In particular, each RB contains two convolutional layers and a skip shortcut, and the FNN includes three fully connected layers. The output of the second RB are flattened and fed into the FNN which will then be used to calculate the corresponding location of the input CSI matrix.


\subsubsection{Training Setup}
Before training the positioning models with the proposed algorithm, we first pretrain the NN at each BS using the labeled dataset $\mathcal{D}_{b}^{\textrm{L}}$ to obtain good initial parameters. During NN initialization, the positioning model at each BS is trained on $\mathcal{D}_{b}^{\textrm{L}}$ for 5000 epochs with a learning rate of $10^{-3}$. This learning rate will be reduced to its 90\% by every 100 epochs when the number of training epochs is larger than 2000. The batch size of gradient descent is set as $B_1 = 32$. Hence, one training epoch will include $\frac{N_{b}^{\textrm{L}}}{B_1}$ iterations. After pretraining, the model at each BS are further trained using the proposed algorithm for 10000 iterations with an initial learning rate of $10^{-3}$. The learning rate will be reduced to its 90\% by every 200 iterations if the number of iterations is larger than 5000. We also select a mini-batch of samples from $\mathcal{D}^{\textrm{M}}$ with a batch size $B_2 = 24$ at each iteration. Subsequently, instead of computing the MSE loss over all $\left\{\mathcal{H}^k , \mathcal{P}^k\right\} \in   \mathcal{D}^{\textrm{M}}$, we only calculate the positioning loss on this mini-batch so as to reduce the training overhead, and the maximization of $\vert \widetilde{\epsilon}_{b,v} \vert$ in (17) will also be conducted on the selected mini-batch rather than on a single data sample $\left\{\mathcal{H}^k , \mathcal{P}^k\right\}$. To maintain the balance between MSE losses on $\mathcal{D}^{\textrm{M}}$ and each $\mathcal{D}_{b}^{\textrm{L}}$, the number of labeled CSI samples at each iteration for computing the loss on $\mathcal{D}_{b}^{\textrm{L}}$ is equal to the number of CSI matrices in the training batch selected from $\mathcal{D}^{\textrm{M}}$.

The specific values of experimental parameters are concluded in Table $\textrm{\uppercase\expandafter{\romannumeral2}}$.

\begin{figure}[t]
\centering
\setlength{\belowcaptionskip}{-0.5cm}
\includegraphics[width=9cm]{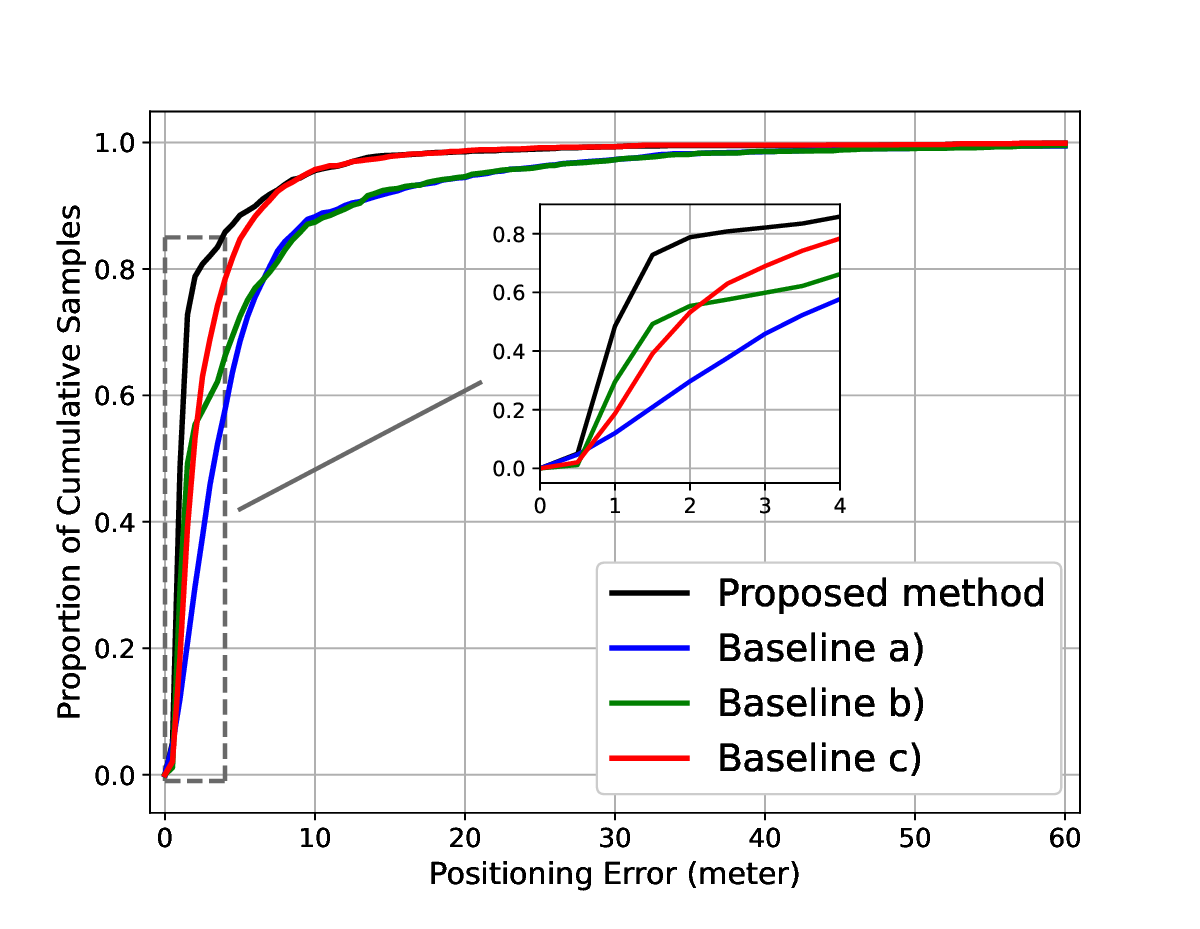}
\centering
\vspace{-0.3cm}
\caption{CDF of positioning errors with $N_{b}^{\textrm{L}}=300$.}
\label{fig6}
\vspace{0.2cm}
\end{figure}


\subsection{Performance Evaluation}
To evaluate the performance of our proposed method, we first calculate the mean Euclidean distances between the predicted and ground truth position coordinates \cite{17,44} at each BS as the positioning error. Then, the average positioning error of all BSs is utilized as the performance measurement.

Fig. \ref{fig6} shows the cumulative distribution functions (CDF) of positioning errors of all the BSs when $N_{b}^{\textrm{L}}=300$. Here, we consider three baselines for comparison purposes. In baseline a), we directly use the models trained with the labeled dataset $\mathcal{D}_{b}^{\textrm{L}}$ to predict the vehicle locations in testing datasets. In baseline b), the models are directly trained with $\mathcal{D}_{b}^{\textrm{L}}$, and the output position coordinates of the NNs will be further calibrated with the locations obtained from images as described in (18). In baseline c), we utilize the proposed algorithm to train the NN at each BS. But the output will not be calibrated in the inference procedure. From this figure, we see that 79\% of positioning errors of the proposed method are smaller than 2m, but only 30\% of positioning errors of baseline a) are lower than 2m. This is because the designed algorithm can effectively train the NN models with multi-modal dataset $\mathcal{D}^{\textrm{M}}$, and the locations predicted by the NNs are calibrated by the position coordinates obtained from images. We also observe from Fig. \ref{fig6} that 55\% of positioning errors of baseline b) are smaller than 2m, which is 24\% inferior to the proposed method. This also proves that the NNs are well trained with the proposed method using $\mathcal{D}^{\textrm{M}}$. Since the NN models in baseline b) are only trained with the labeled dataset $\mathcal{D}_{b}^{\textrm{L}}$, the positioning accuracies of these models are relatively low. The positioning models are effectively trained with $\mathcal{D}^{\textrm{M}}$ in the proposed algorithm, which improves their positioning accuracies, thus decreasing the possibility of incorrect matching between predicted locations and locations obtained from images. Finally, we see that the percentage of positioning errors which are less than 2m of the proposed method is 26\% larger than that of baseline c). This is because most of the predicted locations are correctly matched to the vehicles in images through the proposed inference procedure.

\begin{figure}[t]
\centering
\setlength{\belowcaptionskip}{-0.5cm}
\includegraphics[width=9cm]{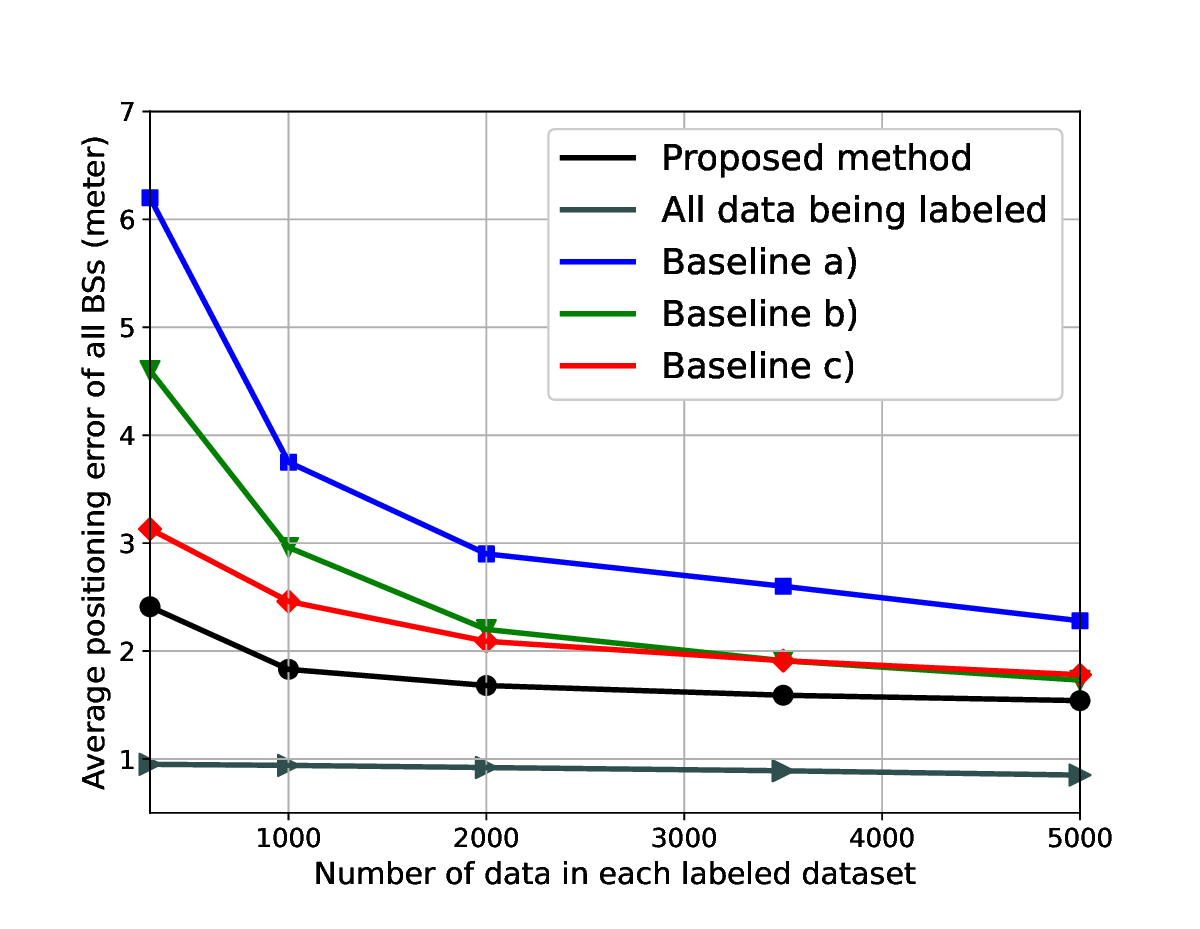}
\centering
\vspace{-0.3cm}
\caption{Average positioning error with different amount of labeled data.}
\label{fig7}
\vspace{0.1cm}
\end{figure}

\begin{figure}[t]
\centering
\setlength{\belowcaptionskip}{-0.5cm}
\includegraphics[width=9cm]{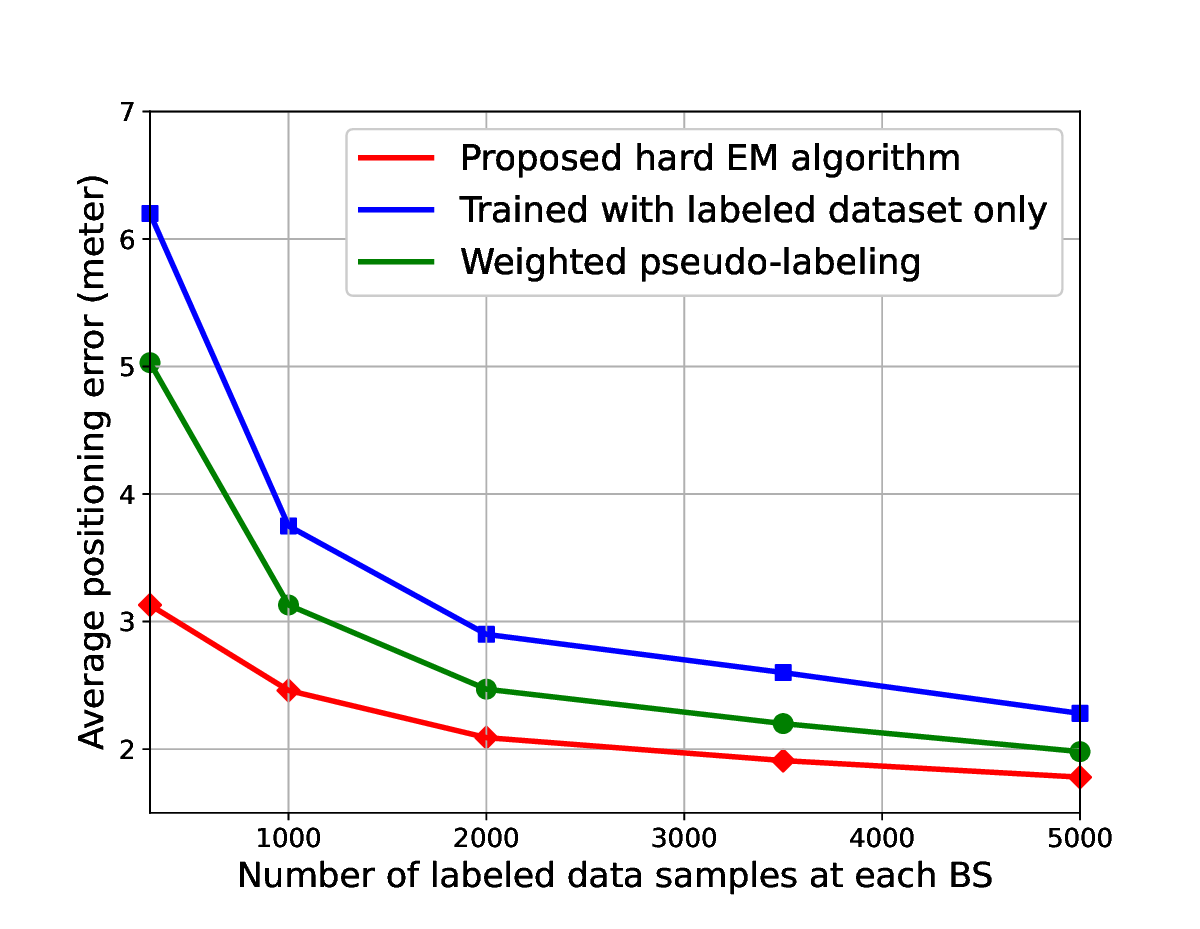}
\centering
\vspace{-0.3cm}
\caption{Positioning error compared to weighted pseudo-labeled technique.}
\label{fig13}
\vspace{0.5cm}
\end{figure}

\begin{figure}[t]
\centering
\setlength{\belowcaptionskip}{-0.5cm}
\includegraphics[width=9cm]{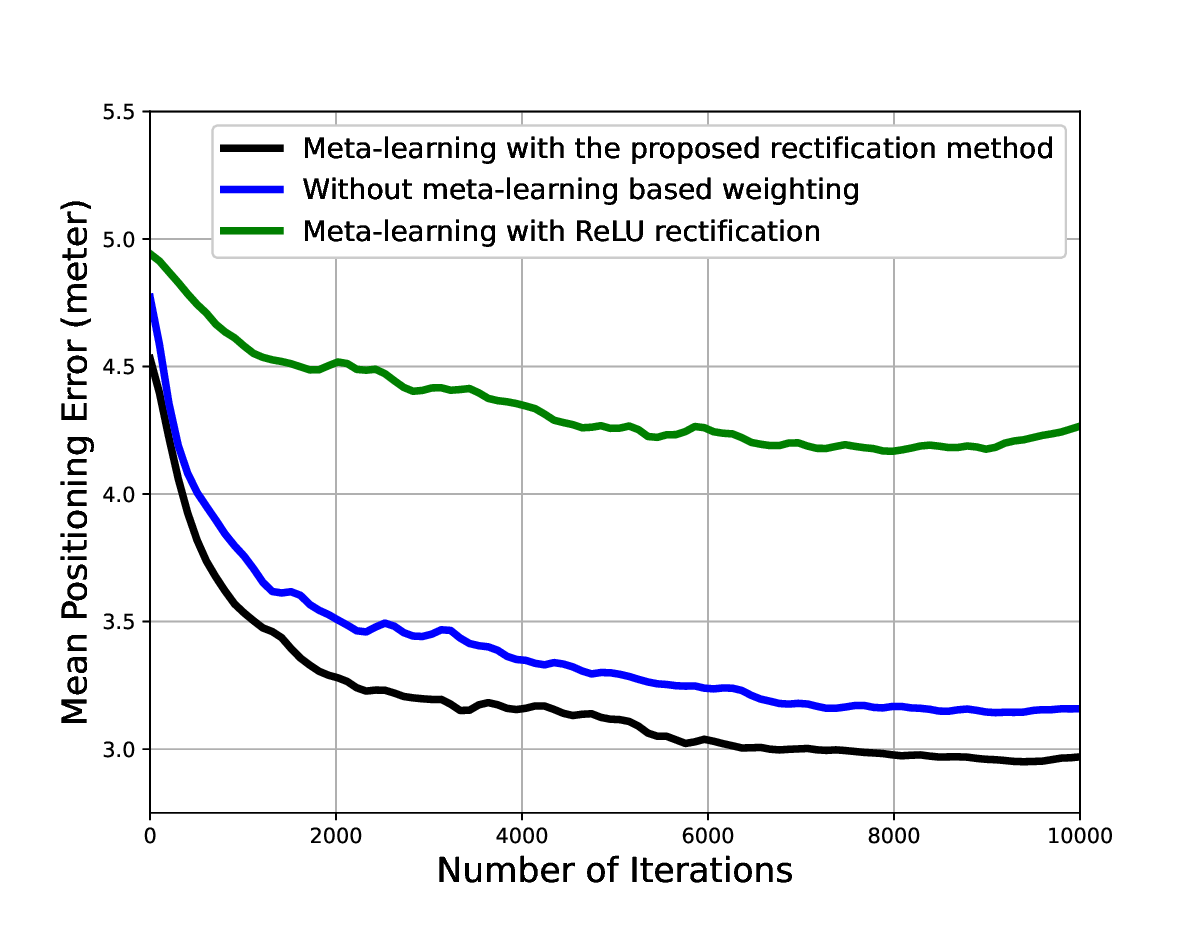}
\centering
\vspace{-0.3cm}
\caption{Testing positioning error with $N_{b}^{\textrm{L}}=300$ using different meta-learning settings.}
\label{fig8}
\vspace{-0.1cm}
\end{figure}

\begin{figure}[t]
\centering
\setlength{\belowcaptionskip}{-0.5cm}
\includegraphics[width=9cm]{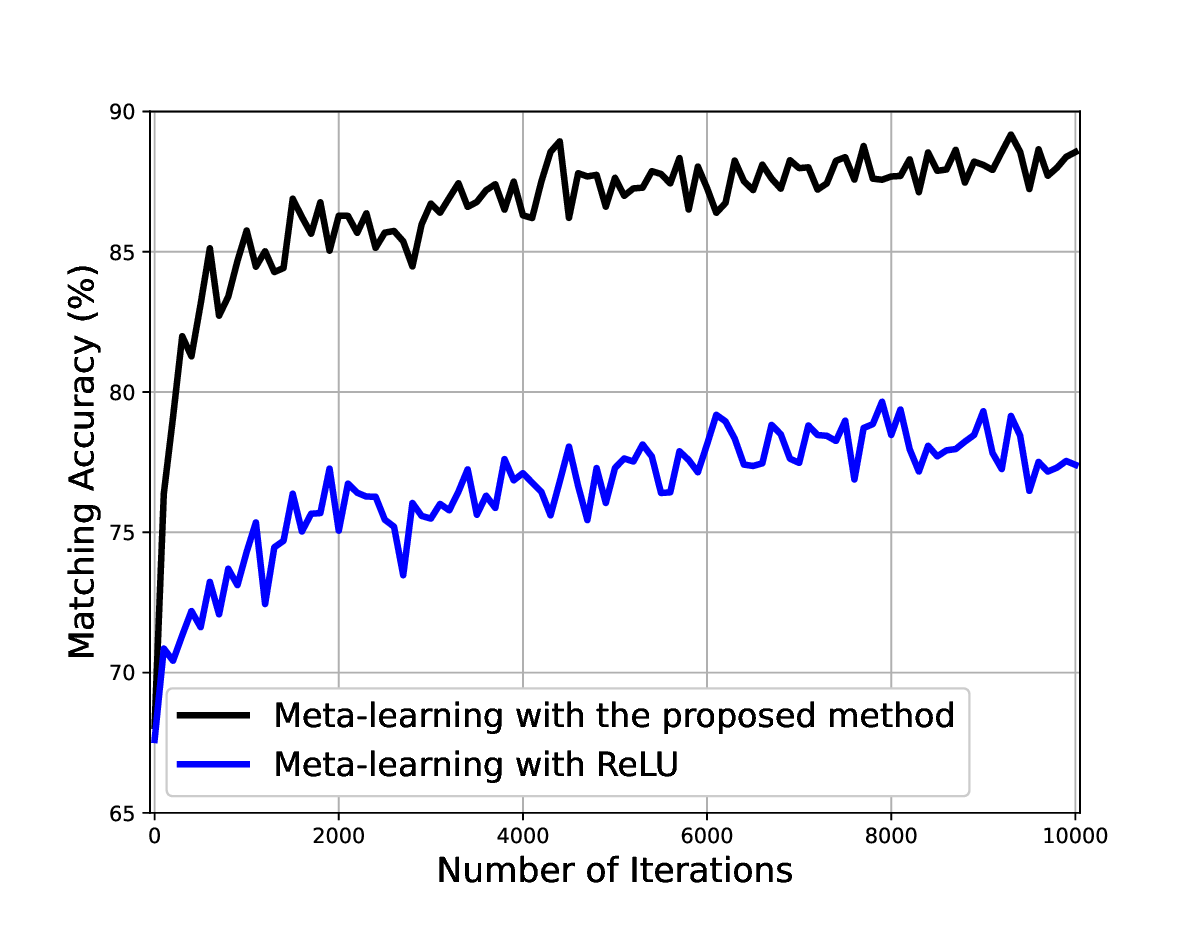}
\centering
\vspace{-0.3cm}
\caption{Matching accuracy with $N_{b}^{\textrm{L}}=300$ using different meta-learning settings.}
\label{fig9}
\vspace{-0.1cm}
\end{figure}

\begin{figure}[t]
\centering
\setlength{\belowcaptionskip}{-0.5cm}
\includegraphics[width=9cm]{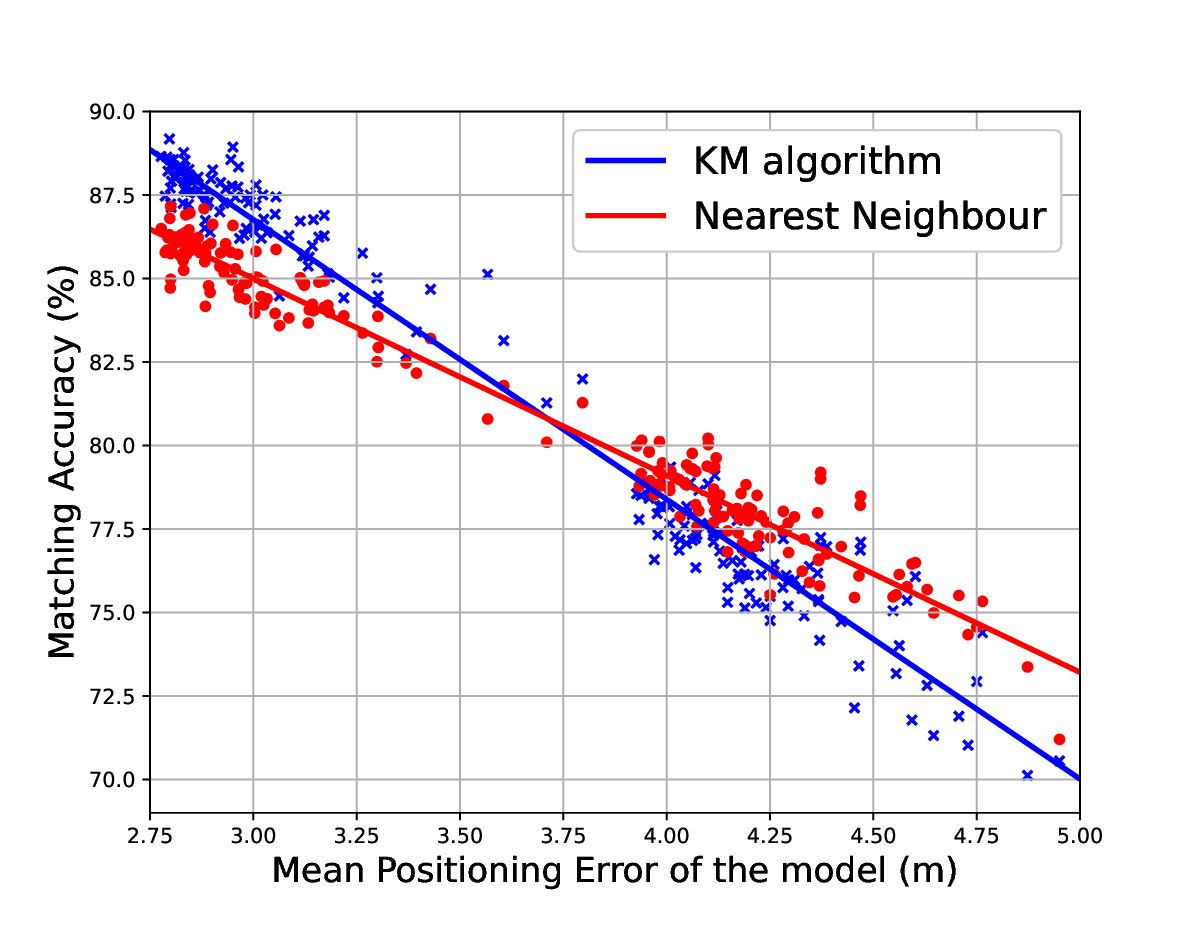}
\centering
\vspace{-0.3cm}
\caption{Relationship between matching accuracy and positioning error using different matching strategies.}
\label{fig10}
\vspace{-0.1cm}
\end{figure}

In Fig. \ref{fig7}, we show how the average positioning error changes as the number $N_{b}^{\textrm{L}}$ of labeled data samples owned by each BS varies. In particular, we consider the same three baselines as in Fig. \ref{fig6}, and we also show the positioning errors of the model which is trained with all the CSI samples in $\mathcal{D}^{\textrm{M}}$ being labeled. From this figure, we see that the average positioning errors of the proposed method are consistently lower than that of the three baselines when $N_{b}^{\textrm{L}}$ changes. We also observe from Fig. \ref{fig7} that as the number of labeled training samples increases, the positioning errors of all considered algorithms decrease. This stems from the fact that the model trained with more labeled samples will have a better generalization capacity, and hence can achieve lower positioning error. However, we observe that when $N_{b}^{\textrm{L}}\geq 2000$, the average positioning error of the proposed method remains the same. This is because most of the predicted locations have been correctly matched to the vehicles in images when $N_{b}^{\textrm{L}}$ is larger than $2000$. As a result, the average positioning error will no longer be determined by the localization accuracy of the NNs, but will be constrained by the distance errors between ground truth locations and the calculated vehicle locations from images.

In Fig. \ref{fig13}, we compare the performance of our proposed method to the weighted pseudo-labeling algorithm \cite{PseudoLabel}, in order to show the advantage of using images to generate labels for CSI samples. In particular, the weighted pseudo-labeling algorithm first uses the model trained with the small labeled dataset to generate pseudo labels for the unlabeled CSI samples, and then uses both labeled and unlabeled CSI samples to retrain the positioning model. From this figure, we see that 
the proposed algorithm can reduce the positioning error by up to 38\% when the number of labeled data is 300, compared to the weighted pseudo-labeling method. This stems from the fact that the accuracy of the label provided by images is much higher than the accuracy of the pseudo label, since the pseudo labels are generated by the model trained only with a small labeled dataset. 

In Fig. \ref{fig8}, we present the mean positioning error of a BS on testing dataset with different meta-learning setups when $N_{b}^{\textrm{L}}=300$. In particular, we consider two baselines to show the effectiveness of the designed meta-learning based weighting algorithm. In the first baseline, the positioning model is trained by using the hard EM algorithm but without meta-learning based weighting. In the second baseline, we rectify the weight parameters with the ReLU method \cite{38} in order to show the superiority of our proposed method in (17). We see from this figure that the designed weighting algorithm can reduce the positioning error by up to 7\% compared to the baseline in which the model is trained without weighting. This is because the incorrect matching between the position labels obtained from images and CSI will introduce noisy training labels. The designed meta-learning based weighting algorithm can reduce the impact of label noises on model training by assigning smaller weight parameters to the data samples with incorrect positioning labels. From Fig. \ref{fig8}, we can also observe that the designed method can reduce the positioning error by up to 29\% compared to the ReLU method. This stems from the fact that the ReLU method will directly discard all the training samples whose gradient directions are opposite to the gradient direction provided by validation dataset $\mathcal{D}_b^{\textrm{V}}$, and thus may discard a large number of noiseless training samples and introduce overfitting on $\mathcal{D}_b^{\textrm{V}}$. Different from ReLU, the proposed method assigns a smaller weight parameter (i.e., $\widetilde{\epsilon}_{b,v}<0$) to the data samples rather than directly discarding these samples during training.

\begin{figure}[t]
\centering
\setlength{\belowcaptionskip}{-0.5cm}
\includegraphics[width=8.7cm]{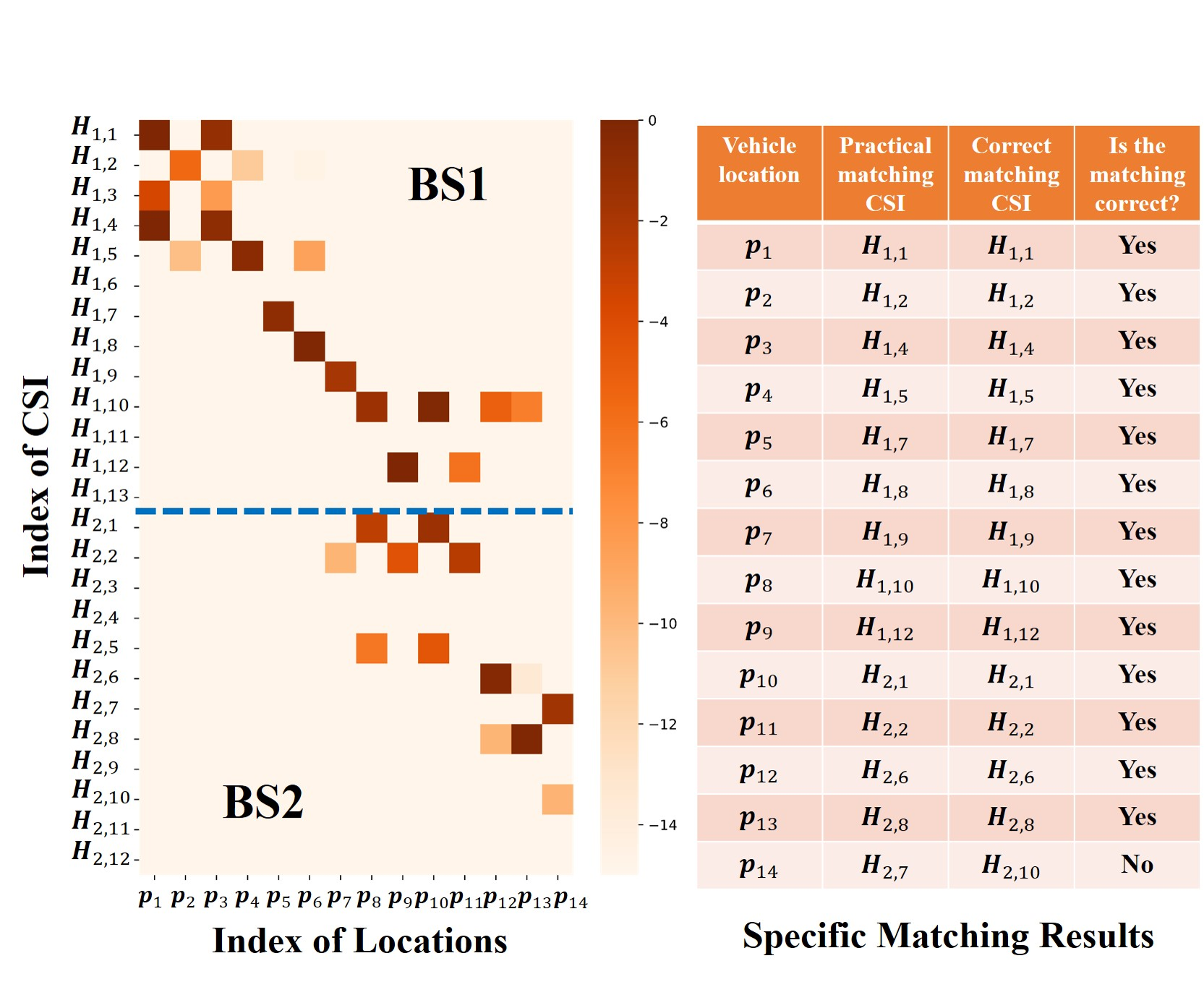}
\centering
\vspace{-0.2cm}
\caption{An example of the inference procedure.}
\label{fig11}
\vspace{0.1cm}
\end{figure}

Fig. \ref{fig9} shows the percentage of the vehicle locations obtained from images which are correctly matched to their corresponding CSI samples when using ReLU and the proposed method in (17). From this figure, we see that the proposed method can increase the matching accuracy by up to 10\% compared to the ReLU method. This is because the model trained using the proposed method has smaller positioning error than the ReLU method. Since the matching accuracy is negatively correlated to the positioning error of the model, the proposed method can achieve a higher matching accuracy within the same number of training iterations.

In Fig. \ref{fig10}, we show the functional relationship between mean positioning error of the positioning model and the percentage of correct matching between unlabeled CSI and vehicles in images when we use different matching strategies. In particular, except for KM algorithm, we consider to use the nearest neighbour algorithm for matching CSI and vehicles in images, where a CSI sample will be matched with the vehicle location in images which is closest to the position of the CSI predicted by NN. From Fig. \ref{fig10}, we observe that when the positioning error is larger than 3.7 m, the matching accuracy of the nearest neighbour algorithm is larger than the KM algorithm. This stems from the fact that the solution of KM algorithm is based on the constraint that each vehicle can only correspond to one unlabeled CSI. Hence, if a vehicle in image is matched to a wrong CSI because of the high positioning error of the model, it can longer be matched with the correct CSI. However, in the nearest neighbour algorithm, the vehicles in images can still be matched to the correct CSI as long as the vehicle locations are closest to the predicted location of the correct CSI. We also see from this figure that the KM algorithm has higher matching accuracy when the positioning error is lower than 3.7 m. This is because the KM algorithm can accurately match unlabeled CSI samples to the correct vehicles in images when the positioning error of the model is low. However, although the model has low positioning error, the nearest neighbour algorithm may still generate wrong matching as long as the nearest vehicle location of an unlabeled CSI is not the correct one.

Fig. \ref{fig11} shows an example of the reward matrix and matching results of the inference procedure with $N_{b}^{\textrm{L}}=300$. Specifically, we present the reward matrix of a testing sample $\left\{\mathcal{H} , \mathcal{P}\right\}$ in the form of a heat map, in which the depth of color is used to represent the reward value of matching a CSI sample to a vehicle location in images. Note that we only show the reward values in the range of $\left[-15,0\right]$ using different colors, and the reward values smaller than $-15$ will be all represented by the same color as $-15$. From this figure, we see that the designed inference procedure can accurately infer the correct corresponding relationship between CSI and vehicles.

\begin{figure}[t]
\centering
\setlength{\belowcaptionskip}{-0.5cm}
\includegraphics[width=8.5cm]{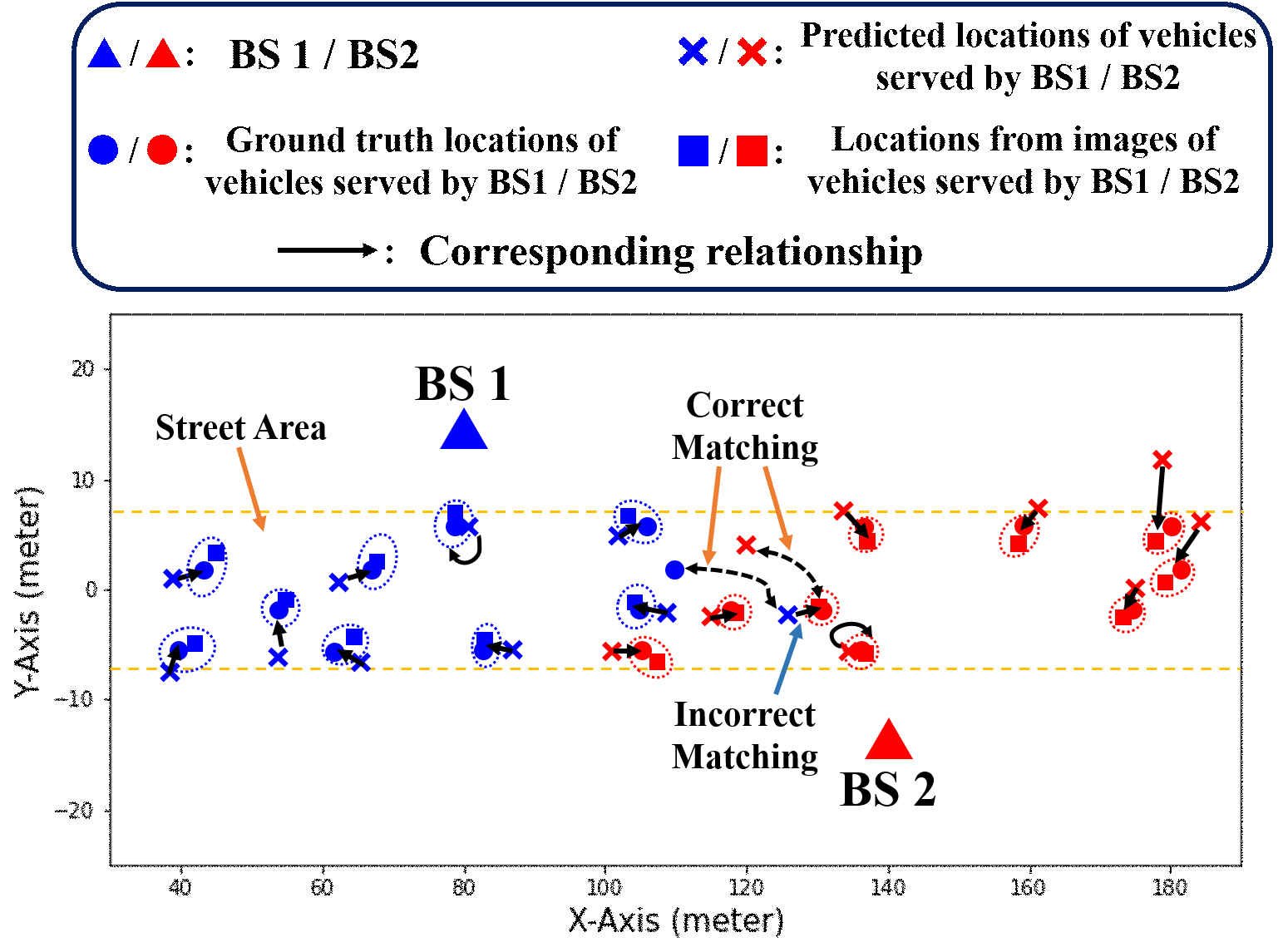}
\centering
\vspace{-0.1cm}
\caption{An example of image aided vehicle positioning.}
\label{fig12}
\vspace{0.1cm}
\end{figure}

Fig. \ref{fig12} shows an example of using images to assist vehicle positioning when $N_{b}^{\textrm{L}}=1000$. From this figure, we see that most of the predicted vehicle locations are correctly matched to the locations obtained from images. Hence, compared to the method that localizes vehicles using only CSI data, the mean positioning error of the proposed algorithm that uses multi-modal data is much lower.

\section{Conclusion}
In this paper, we have developed a novel multi-modal multi-BS vehicle positioning framework that jointly utilize images and CSI fingerprints to localize vehicles. We have considered a practical communication scenario where each vehicle can communicate with only one BS at the same time, and hence, it can upload its estimated CSI to only its associated BS. Each BS is equipped with a set of cameras, and it can collect a large number of unlabeled CSI, the images taken by cameras, and only a small amount of labeled CSI for model training. To exploit the unlabeled CSI data and position labels obtained from images, we design a meta-learning based hard EM algorithm that iteratively trains the positioning model of each BS. Simulation results show that the proposed method can outperform the baseline where the images are not used to aid the CSI fingerprint based vehicle positioning.

\appendix
\subsection{Proof of Lemma 1}
To prove lemma 1, we first rewrite the gradient difference $\boldsymbol{\omega}_b^{i+1} - \boldsymbol{\omega}_b^{i}$ on $\left\{\mathcal{H}^k,\hat{\mathcal{R}}_{\boldsymbol{\Omega}^{i-1}}^{k}\right\}$ with $\gamma=1$, which can be expressed as
\begin{equation}
\begin{aligned}
\boldsymbol{\omega}_b^{i+1} - \boldsymbol{\omega}_b^{i} &= -\lambda_{i} \nabla \frac{\sum_{v=1}^{V_b} 
\| \hat{\boldsymbol{r}}_{b,v}\left(\hat{\boldsymbol{\alpha}}_{b,v}\right) 
- F_{\boldsymbol{\omega}}\big(\boldsymbol{H}_{b,v}\big)\|_2^2}{
\sum_{v=1}^{V_b} \big(1 - \hat{\alpha}_{b,v}^{\hat{V} +1}\big)}\\
&=-\lambda_{i} \nabla \Big[\frac{1}{n_b} \sum_{v=1}^{V_b} \hat{\epsilon}_{b,v} \cdot
 g_{b,v}\left(\boldsymbol{\omega}_b^{i}\right)\Big].
\end{aligned}\tag{22}
\end{equation}
Then, substituting (22) into (20), we can obtain
\begin{equation}
\begin{aligned}
&G_b \left(\boldsymbol{\omega}_b^{i+1}\right) - G_b \left(\boldsymbol{\omega}_b^{i}\right) 
\\
&\;\;\;\;\;\;\;\;\;\;\;\; \leq -\frac{\lambda_{i}}{n_b}\sum_{v=1}^{V_b}\hat{\epsilon}_{b,v} \Big\langle \nabla G_b \left(\boldsymbol{\omega}_b^{i}\right) , \nabla g_{b,v}\left(\boldsymbol{\omega}_b^{i}\right) \Big\rangle
\\ 
&\;\;\;\;\;\;\;\;\;\;\;\;\;\;\;\;\;\;\;\;\;\;\;\; +\frac{L \lambda_i^2}{2 n_b^2} \Big[\sum_{v=1}^{V_b} \hat{\epsilon}_{b,v} \cdot \nabla
 g_{b,v}\left(\boldsymbol{\omega}_b^{i}\right)\Big]^2 .
\end{aligned}\tag{23}
\end{equation}
Using the triangle inequality and the upper bound $\beta$ of $\|\nabla
 g_{b,v}\left(\boldsymbol{\omega}_b^{i}\right) \|_2$, we have 
\begin{equation}
\begin{aligned}
\Big[\sum_{v=1}^{V_b} \hat{\epsilon}_{b,v}  \nabla
 g_{b,v}\left(\boldsymbol{\omega}_b^{i}\right)\Big]^2 \!\!\! \leq \sum_{v=1}^{V_b} \hat{\epsilon}_{b,v}^2  \|  \nabla
 g_{b,v}\left(\boldsymbol{\omega}_b^{i}\right)  \|_2^2
 \leq  \sum_{v=1}^{V_b} \hat{\epsilon}_{b,v}^2 \beta^2 .
\end{aligned}\tag{24}
\end{equation}
Hence, based on (23) and (24), we obtain
\begin{equation}
\begin{aligned}
&G_b \left(\boldsymbol{\omega}_b^{i+1}\right) - G_b \left(\boldsymbol{\omega}_b^{i}\right) 
\\
&\leq -\frac{\lambda_{i}}{n_b}\sum_{v=1}^{V_b}\Big[\hat{\epsilon}_{b,v} \Big\langle \nabla G_b \left(\boldsymbol{\omega}_b^{i}\right) , \nabla g_{b,v}\left(\boldsymbol{\omega}_b^{i}\right) \Big\rangle -\frac{L  \beta^2 \lambda_i}{2 n_b}\hat{\epsilon}_{b,v}^2 \Big]. 
\end{aligned}\tag{25}
\end{equation}
Since we assume that $\xi=1$, according to (16) and (17), we have
\begin{equation}
\Big\langle \nabla G_b \left(\boldsymbol{\omega}_b^{i}\right) , \nabla g_{b,v}\left(\boldsymbol{\omega}_b^{i}\right) \Big\rangle = \mu_b \left(\hat{\epsilon}_{b,v}-1 \right),\tag{26}
\end{equation}
where $\mu_b=\mathop{\max}\limits_{v}\vert \widetilde{\epsilon}_{b,v} \vert$. Then, substituting (26) into (25), we can obtain
\begin{equation}
\begin{aligned}
&G_b \left(\boldsymbol{\omega}_b^{i+1}\right) - G_b \left(\boldsymbol{\omega}_b^{i}\right) 
\\
&\leq -\frac{\lambda_{i}}{n_b}\sum_{v=1}^{V_b}\Big[ \mu_b \hat{\epsilon}_{b,v} \left(\hat{\epsilon}_{b,v}-1 \right) -\frac{L  \beta^2 \lambda_i}{2 n_b}\hat{\epsilon}_{b,v}^2\Big]
\\
&\approx -\frac{\lambda_{i} V_b}{n_b} \cdot \mathbb{E}_{\epsilon}\Big[  \Big(\mu_b-\frac{L  \beta^2 \lambda_i}{2 n_b} \Big)\epsilon^2 -\mu_b \epsilon\Big].
\end{aligned}\tag{27}
\end{equation}
Since we assume that $\widetilde{\epsilon}_{b,v} \sim U\left(-\mu_b,\mu_b\right)$, according to (17), we have $\epsilon \sim U\left(1-\xi,1+\xi\right)$, i.e. $\epsilon \sim U\left(0,2\right)$ when given $\xi=1$. Therefore, we have
\begin{equation}
\begin{aligned}
&G_b \left(\boldsymbol{\omega}_b^{i+1}\right) - G_b \left(\boldsymbol{\omega}_b^{i}\right) 
\\
&\leq  -\frac{\lambda_{i} V_b}{n_b} \int_{0}^{2} \frac{1}{2}\Big[\Big(\mu_b-\frac{L  \beta^2 \lambda_i}{2 n_b} \Big)\epsilon^2 -\mu_b \epsilon \Big]\; \mathrm{d} \epsilon
\\
&= -\frac{\lambda_{i} V_b}{n_b} \cdot \frac{1}{3} \Big(\mu_b-\frac{ 2L  \beta^2 \lambda_i}{n_b} \Big).
\end{aligned}\tag{28}
\end{equation}
Hence, if $\mu_b-\frac{ 2L  \beta^2 \lambda_i}{n_b} \geq 0$, namely $\lambda_i\leq \frac{\mu_b n_b }{2L \beta^2}$, the validation loss is guaranteed to be monotonically non-increasing.

Next, we prove that $\mathbb{E}\left[G_b \left(\boldsymbol{\omega}_b^{i+1}\right)-G_b \left(\boldsymbol{\omega}_b^{i}\right)\right]=0$, if and only if $\nabla G_b \left(\boldsymbol{\omega}_b^{i}\right)=0$. Here, we assume that $\lambda_i = \delta \cdot \frac{\mu_b n_b }{2L \beta^2}$, where $\delta \in \left(0,1\right)$ is a scale parameter, in order to ensure that $G_b \left(\boldsymbol{\omega}\right)$ is non-increasing. Then, based on (28), we have
\begin{equation}
\mathbb{E}\Big[G_b \left(\boldsymbol{\omega}_b^{i+1}\right) - G_b \left(\boldsymbol{\omega}_b^{i}\right)\Big] \leq
- \frac{\delta \big(1 - \delta\big)V_b}{6 L \beta^2}\, \mathbb{E}\big[\mu_b^2\big].
\tag{29}
\end{equation}
Hence, if $\mathbb{E}\left[G_b \left(\boldsymbol{\omega}_b^{i+1}\right)-G_b \left(\boldsymbol{\omega}_b^{i}\right)\right]=0$, we have $\mu_b^2 =0$, i.e. $\mathop{\max}\limits_{v}\Big\langle \nabla G_b \left(\boldsymbol{\omega}_b^{i}\right) , \nabla g_{b,v}\left(\boldsymbol{\omega}_b^{i}\right) \Big\rangle^2 =0$. Therefore, it must hold that $\nabla G_b \left(\boldsymbol{\omega}_b^{i}\right) = 0$. On the other hand, if $\nabla G_b \left(\boldsymbol{\omega}_b^{i}\right) = 0$, we have $\mu_b= 0$. Hence, it holds that $\lambda_i = \delta \cdot \frac{\mu_b n_b }{2L \beta^2} = 0$. As a result, the updated parameter $\boldsymbol{\omega}_b^{i+1}$ is equal to $\boldsymbol{\omega}_b^{i}$, which implies that $G_b \left(\boldsymbol{\omega}_b^{i+1}\right)-G_b \left(\boldsymbol{\omega}_b^{i}\right)=0$. 

This completes the proof.
\hfill $\Box$

\bibliographystyle{IEEEtran}
\bibliography{myrefs}


\end{document}